\documentclass[aps, prl, floatfix,
  reprint, 
  superscriptaddress, 
  longbibliography, noeprint, nofootinbib
]{revtex4-2} 
\usepackage{graphicx, tikz, xcolor} 
\usepackage[notrig]{physics} 
\usepackage{amsmath, amssymb}
\usepackage{mathtools} 
\usepackage[pdftex,unicode,pdfusetitle]{hyperref}
\usepackage{natbib, hypernat}
\usepackage[resetlabels]{multibib}
\newcites{SM}{ }
\hypersetup{
	backref,
	colorlinks=true,
	citecolor=blue,
    linkcolor=magenta,
    urlcolor=blue,
    breaklinks=true,
	pdfborder={0 0 0},
	pdfauthor={Shoki Sugimoto, Ryusuke Hamazaki, and Masahito Ueda},
}
\graphicspath{ {./Figure/} }

\makeatletter
\def\frontmatter@maketitle{%
  \@author@finish
  \title@column\titleblock@produce
  \suppressfloats[t]%
  \let\abstract\@undefined\let\endabstract\@undefined
  \titlepage@sw{%
   \vfil
   \clearpage
  }{}%
  \onecolumn@grid@setup
  \def\set@footnotewidth{\set@footnotewidth@one}%
}%
\makeatother

\begin{document}
\title{
  Eigenstate Thermalization in Long-Range Interacting Systems
}
\author{Shoki Sugimoto}
  \affiliation{Department of Physics, The University of Tokyo, 7-3-1 Hongo, Bunkyo-ku, Tokyo 113-0033, Japan}
\author{Ryusuke Hamazaki}
  \affiliation{Nonequilibrium Quantum Statistical Mechanics RIKEN Hakubi Research Team, RIKEN Cluster for Pioneering Research (CPR), RIKEN iTHEMS, Wako, Saitama 351-0198, Japan}
\author{Masahito Ueda}
  \affiliation{Department of Physics, The University of Tokyo, 7-3-1 Hongo, Bunkyo-ku, Tokyo 113-0033, Japan}
  \affiliation{RIKEN Center for Emergent Matter Science (CEMS), Wako 351-0198, Japan}
\begin{abstract}
    Motivated by recent ion experiments on tunable long-range interacting quantum systems~[B.~Neyenhuis \textit{et al.}, \href{https://doi.org/10.1126/sciadv.1700672}{Sci.~Adv.~\textbf{3}, e1700672 (2017)}], we test the strong eigenstate thermalization hypothesis~(ETH) for systems with power-law interactions $\sim 1/r^{\alpha}$.
    We numerically demonstrate that the strong ETH typically holds at least for systems with $\alpha\geq 0.6$, which include Coulomb, monopole-dipole, and dipole-dipole interactions.
    Compared with short-range interacting systems, the eigenstate expectation value of a generic local observable is shown to deviate significantly from its microcanonical ensemble average for long-range interacting systems.
    We find that Srednicki's ansatz breaks down for $\alpha \lesssim 1.0$ at least for relatively large system sizes.
\end{abstract}
\maketitle

\paragraph{Introduction.---}  
Long-range interacting systems show a number of unique phenomena~\cite{dauxois2002dynamics, campa2009statistical, campa2014physics, defenu2021longrange} such as negative heat capacity~\cite{schmidt2001negative, gobet2001probing}, anomalous propagation of correlations~\cite{hauke2013spread,schachenmayer2013entanglement,richerme2014non,jurcevic2014quasiparticle, cevolani2018universal, schneider2021spreading}, and prethermalization~\cite{van2013relaxation, gong2013prethermalization, marcuzzi2013prethermalization, mori2019prethermalization, defenu2021metastability}.
Isolated quantum systems with long-range interactions have been realized in trapped ion systems~\cite{blatt2012quantum}, Rydberg atom arrays~\cite{browaeys2020many} and quantum gases coupled to optical cavities~\cite{RevModPhys.85.553}.
The dynamic~\cite{gong2013prethermalization, van2013relaxation,smith2016many,neyenhuis2017observation} and thermodynamic~\cite{islam2011onset, islam2013emergence, richerme2014non} properties of these systems have also been investigated.
In particular, trapped ion systems offer an ideal platform for the study of isolated quantum systems with long-range interactions $\sim 1/r^{\alpha}$, where the exponent $\alpha$ can be tuned from $0$ to $3$ by a spin-dependent optical dipole force~\cite{porras2004effective, kim2009entanglement, islam2011onset, britton2012engineered, yoshimura2015creation, bohnet2016quantum, richerme2016two, hess2017non}.

Prethermalization of a long-range nonintegrable quantum system without disorder was experimentally observed~\cite{neyenhuis2017observation}, but complete thermalization was not observed in an experimentally accessible time.
This appears inconsistent with the strong eigenstate thermalization hypothesis~(ETH)~\cite{neumann1929beweis, deutsch1991quantum, srednicki1994chaos}
which states that an expectation value $O_{\gamma\gamma}$ of a physical observable $\hat{O}$ for \textit{every} energy eigenstate $\ket*{E_{\gamma}}$ of a quantum many-body Hamiltonian agrees with its microcanonical ensemble average in the thermodynamic limit~\cite{rigol2008thermalization, rigol2009breakdown, biroli2010effect, khatami2012quantum, polkovnikov2011colloquium, eisert2015quantum,gogolin2016equilibration, d2016quantum, mori2018thermalization, deutsch2018eigenstate}.
We formulate this statement as~\cite{sugimoto2021test}
\begin{equation}
    \Delta_{\infty} \coloneqq \frac{ \max\abs{ O_{\gamma\gamma} - \expval*{\hat{O}}_{\delta E}^{\mathrm{mc}}(E_{\gamma}) } }{ \eta_{O} } \xrightarrow{N\to\infty} 0, \label{Eq:StrongETH}
\end{equation}
where $\eta_{O}$ is the spectral range of $\hat{O}$ defined as the difference between the maximum and minimum eigenvalues of $\hat{O}$, and $\expval*{\hat{O}}_{\delta E}^{\mathrm{mc}}(E_{\gamma})$ is the microcanonical average of $\hat{O}$ in an energy shell $\mathcal{H}_{E_{\gamma},\delta E}$ centered at $E_{\gamma}$ with a sufficiently small width $2\delta E$.
The strong ETH has numerically been verified to hold for various short-range interacting systems~\cite{rigol2010quantum, santos2010localization,PhysRevE.87.012118,beugeling2014finite,kim2014testing,steinigeweg2014pushing, jansen2019eigenstate, sugimoto2021test}.
However, little is known about the validity of the strong ETH in long-range interacting systems except for a few specific models~\cite{khatami2012quantum, khatami2013fluctuation, mori2017classical}.

In this Letter, we test the typicality of the strong ETH for spin systems with power-law interactions $\sim 1/r^{\alpha}$ by introducing an \textit{ensemble} of such systems.
Our result is based on numerical diagonalization, since analytically addressing the strong ETH is extremely difficult because of a chaotic nature of energy eigenstates satisfying the ETH~\cite{berry1977regular,srednicki1994chaos} and the few-body constraint of realistic operators.
We find that the strong ETH typically holds at least for $\alpha \geq 0.6$ in one dimension.
For $\alpha \leq 0.5$, we find no evidence in support of the strong ETH for system size up to 20 spins relevant to trapped-ion experiments~\cite{islam2011onset, richerme2014non, jurcevic2014quasiparticle, neyenhuis2017observation}.
We also test Srednicki's ansatz~\cite{srednicki1999approach}, which states that (i) the deviation $\delta O_{\gamma\gamma} \coloneqq O_{\gamma\gamma} -\expval*{ \hat{O} }_{\delta E}^{\mathrm{mc}}(E_{\gamma})$ behaves like a random variable satisfying
\begin{equation}
    \mathcal{E}[\delta O_{\gamma\gamma}] = 0\qc \mathcal{S}[\delta O_{\gamma\gamma}] = e^{ -\frac{ S(E_{\gamma}) }{2} } f( E_{\gamma} ), \label{eq:SrednickiAnsatz}
\end{equation}
where $\mathcal{E}$ and $\mathcal{S}$ denote the mean and the standard deviation, respectively, $S$ is the thermodynamic entropy, $f$ is a smooth function, and (ii) the distribution of $\delta O_{\gamma\gamma}$ is \textit{Gaussian}~\cite{beugeling2014finite, beugeling2015off, chandran2016eigenstate, mondaini2017eigenstate, lan2017eigenstate,PhysRevE.99.042116, jansen2019eigenstate}.
We find that both (i) and (ii) typically break down for $\alpha \lesssim 1.0$ at least for relatively large system sizes.
These results imply the presence of an intermediate regime $0.5\lesssim\alpha \lesssim 1.0$ in which the strong ETH typically holds, yet Srednicki's ansatz breaks down.

Our results should be distinguished from previous works concerning typical properties of Gaussian random matrices~\cite{neumann1929beweis, goldstein2010long, reimann2015generalization}, banded random matrices~\cite{brandino2012quench, reimann2015eigenstate}, and $k$-body embedded random matrices~\cite{mon1975statistical,kota2001embedded,benet2003review}.
These works do not consider correlations between off-diagonal elements due to interactions, and it is unclear how these correlations affect the typicality of the strong ETH~\cite{hamazaki2018atypicality,sugimoto2021test}.
Our work incorporates such nontrivial correlations by explicitly constructing an ensemble of operators with long-range interactions.

\paragraph{Setup.---}
We consider a one-dimensional spin-1/2 chain of length $N$ subject to the periodic boundary condition.
We denote the local Hilbert space on each site by $\mathcal{H}_{\mathrm{loc}}$ with $d_{L} \coloneqq \dim \mathcal{H}_{\mathrm{loc}} \, (=2)$, the space of all Hermitian operators acting on a Hilbert space $\mathcal{H}$ by $\mathcal{L}(\mathcal{H})$, and an orthonormal basis of $\mathcal{L}(\mathcal{H}_{\mathrm{loc}})$ by $\Bqty*{ \hat{\sigma}^{(p)} }$~\cite{basischoice}.
In numerical calculations, we set $\hat{\sigma}^{(0)} \coloneqq \hat{I}$ and  $\hat{\sigma}^{(p)} \ (p=1,2,3)$ to be the Pauli operators.
For each $\alpha$, $N$, and two-body operator $\hat{h}\in\mathcal{L}(\mathcal{H}_{\mathrm{loc}}^{\otimes 2})$ with $h_{pq} \coloneqq \tr(\hat{h}\, \hat{\sigma}^{(p)}\otimes\hat{\sigma}^{(q)} )/4$, we obtain
\begin{equation}
    \hat{H}_{N}^{(\alpha)}[\hat{h}] \coloneqq \sum_{p,q=1}^{ d_{L}^2-1 } h_{pq} \qty( \sum_{j\neq k}^{N} \frac{ \hat{\sigma}^{(p)}_{j} \hat{\sigma}^{(q)}_{k} }{(r_{jk})^{\alpha}} ), \label{Eq:Hamiltonian}
\end{equation}
where $r_{jk} \coloneqq \min\Bqty*{ \abs*{j-k}, N-\abs*{j-k} }$ is the minimum distance between the sites $j$ and $k$ under periodic boundary condition.
The operator~\eqref{Eq:Hamiltonian} is invariant under translation $\hat{T}_{N} \hat{\sigma}^{(p)}_{j} \hat{T}_{N}^{\dagger} = \hat{\sigma}^{(p)}_{j+1}$ and the parity transformation $\hat{P}_{N}\hat{\sigma}^{(p)}_{j}\hat{P}_{N}^{\dagger} = \hat{\sigma}^{(p)}_{N+1-j}$~\cite{supplement,timeReversalSymmetry}, and does not contain spatially random interactions or random on-site potentials.
In numerical calculations, we focus on the zero-momentum and even-parity sector.

To discuss the typicality of the strong ETH and Srednicki's ansatz, we introduce a set of operators in Eq.~\eqref{Eq:Hamiltonian} by 
$
    \mathcal{G}_{N}^{(\alpha)} \coloneqq \Bqty*{ \hat{H}_{N}^{(\alpha)}[\hat{h}] \mid \hat{h}\in \mathcal{L}(\mathcal{H}_{\mathrm{loc}}^{\otimes 2}) }
$.
The set $\mathcal{G}_{N}^{(\alpha)}$ is quite general as it contains arbitrary two-body long-range operators including Ising, XYZ, Heisenberg models, etc., with arbitrary homogeneous on-site potentials and two-body long-range perturbations.
We sample each $h_{pq}$ in Eq.~\eqref{Eq:Hamiltonian} independently from the standard normal distribution, thereby introducing a probability measure on $\mathcal{G}_{N}^{(\alpha)}$~\cite{measureinvariance}.
For the ensemble of observables, we consider the short-range ensemble $\mathcal{G}_{N}^{(\infty)}$ with only nearest-neighbor and on-site terms.
We investigate the typicality of the strong ETH and Srednicki's ansatz 
by independently sampling Hamiltonians from $\mathcal{G}_{N}^{(\alpha)}$ and observables from $\mathcal{G}_{N}^{(\infty)}$.
\paragraph{Finite-size scaling of the strong ETH measure.---}
Because of Markov's inequality, the typicality of the ETH holds 
if the ensemble average $\mathbb{E}_{N}^{(\alpha)}\bqty*{ \Delta_{\infty} }$ of the dimensionless and intensive measure $\Delta_{\infty}$ of the strong ETH defined in Eq.~\eqref{Eq:StrongETH} vanishes in the thermodynamic limit~\cite{sugimoto2021test}.
We numerically investigate the $N$-dependence of $\mathbb{E}_{N}^{(\alpha)}\bqty*{ \Delta_{\infty} }$,
where $\alpha$ ranges from $0$ to $3$.
Figure~\ref{fig:EnsembleAverage}(a) shows that long-range two-body interactions make $\mathbb{E}_{N}^{(\alpha)}\bqty*{ \Delta_{\infty} }$ significantly larger than that for short-range interacting systems and thus disfavor the strong ETH at least for finite-size systems.

To infer the behavior of $\mathbb{E}_{N}^{(\alpha)}\bqty*{ \Delta_{\infty} }$ in the thermodynamic limit, we analyze the $N$-dependence of $\mathbb{E}_{N}^{(\alpha)}\bqty*{ \Delta_{\infty} }$.
For Gaussian random matrices, where the few-bodiness of realistic operators are completely disregarded, the asymptotic $N$-dependence of $\mathbb{E}_{N}\bqty*{ \Delta_{\infty} }$ is obtained as
\begin{equation}
    \mathbb{E}_{N}^{(\mathrm{RMT})}\bqty*{ \Delta_{\infty} }
\simeq C Ne^{ -N/N_{m} } \sqrt{ 1-\frac{N_{m}}{2} \frac{\log N}{N} -\frac{N_{0}}{N} }, \label{Eq:Asymptotic}
\end{equation}
where $C$, $N_{m}$, and $N_{0}$ are constants~\cite{sugimoto2021test}.
Thus, the concave behavior in $N$ is expected for $\mathbb{E}_{N}^{(\alpha)}\bqty*{ \Delta_{\infty} }$, and it is therefore important to check whether numerically obtained $\mathbb{E}_{N}^{(\alpha)}\bqty*{ \Delta_{\infty} }$ decreases for larger $N$~\cite{fitting}.

The level of confidence that $\mathbb{E}_{N}^{(\alpha)}\bqty*{ \Delta_{\infty} }$ 
decreases with increasing $N$ can be measured by the probability of obtaining a sequence of the estimator $\Bqty*{\hat{\mathbb{E}}_{N_\mathrm{min}}^{(\alpha)}\bqty*{ \Delta_{\infty} },\cdots,\hat{\mathbb{E}}_{N_\mathrm{max}}^{(\alpha)}\bqty*{ \Delta_{\infty} }}$ such that $\hat{\mathbb{E}}_{N_{\min}}^{(\alpha)}[\Delta_{\infty}] > \hat{\mathbb{E}}_{N_{\min}+2}^{(\alpha)}[\Delta_{\infty}] > \dots > \hat{\mathbb{E}}_{N_{\max}}^{(\alpha)}[\Delta_{\infty}]$ in bootstrap iterations~(see Supplemental Material~\cite{supplement} for details).
Figure~\ref{fig:EnsembleAverage}(b) shows that
$\mathbb{E}_{N}^{(\alpha)}\bqty*{ \Delta_{\infty} }$ for $\alpha\geq 0.6$ decreases for large $N$~\cite{note_EvenOdd}.
Therefore, the strong ETH typically holds at least for $\alpha\geq 0.6$.
For $\alpha\leq 0.5$, $\mathbb{E}_{N}^{(\alpha)}\bqty*{ \Delta_{\infty} }$ does not decrease within statistical errors.
While this result suggests the breakdown of the strong ETH for $\alpha\leq0.5$,  we cannot exclude the possibility that $\mathbb{E}_{N}^{(\alpha)}\bqty*{ \Delta_{\infty} }$ vanishes in the thermodynamic limit and hence the strong ETH typically holds for $0<\alpha\leq 0.5$.
Nevertheless, our results for finite-size systems are relevant to trapped-ion experiments~\cite{islam2011onset, richerme2014non, jurcevic2014quasiparticle, neyenhuis2017observation}, where systems involve several tens of ions.
For the fully connected case ($\alpha=0$), the strong ETH typically breaks down in arbitrary dimensions because permutation operators of any two neighboring sites are conserved.
This result is consistent with a monotonically increasing behavior of $\mathbb{E}_{N}^{(\alpha)}[\Delta_{\infty}]$ for $\alpha\simeq 0$ in Fig.~\ref{fig:EnsembleAverage}.

\begin{figure}[tb]
    \centering
    \includegraphics[width=\linewidth]{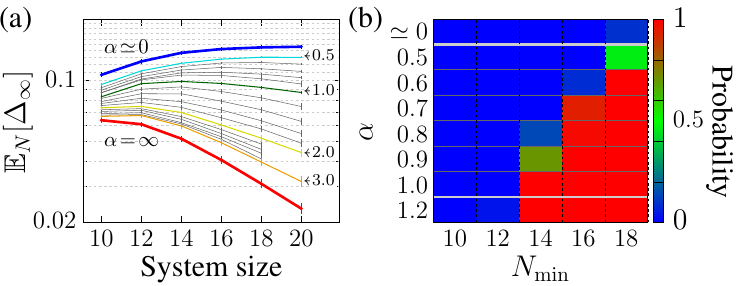}
    \caption{\label{fig:EnsembleAverage}
    (a) Ensemble-averaged strong ETH measure $\Delta_{\infty}$ in Eq.~\eqref{Eq:StrongETH} for tunable-range interactions~$\sim 1/r^{\alpha}$.
    To break the degeneracy in the fully connected case~$(\alpha=0)$, we set $\alpha=0.0001$, which is small enough to capture the essential physics for $\alpha=0$.
    Thin curves between $\alpha=0.5$ and $\alpha=1$ show the data for $\alpha=0.6,0.7,0.8,0.9$, and those between $\alpha=1$ and $\alpha=3$ are for $\alpha=1.2,1.4,\dots,2.8$.
    Each error bar shows the $80\%$ confidence interval.
    (b) Probability of obtaining a sequence $\Bqty*{ \hat{\mathbb{E}}_{N}^{(\alpha)}[\Delta_{\infty}] }_{N_{\min}}^{N_{\max}}$ of the estimator for $\mathbb{E}_{N}^{(\alpha)}[\Delta_{\infty}]$ such that $\hat{\mathbb{E}}_{N_{\min}}^{(\alpha)}[\Delta_{\infty}] > \hat{\mathbb{E}}_{N_{\min}+2}^{(\alpha)}[\Delta_{\infty}] > \dots > \hat{\mathbb{E}}_{N_{\max}}^{(\alpha)}[\Delta_{\infty}]$ with $N_{\max} = 20$, represented by the color of the cell.
    This result shows that the average $\mathbb{E}_{N}^{(\alpha)}[\Delta_\infty]$ decreases for $\alpha\geq 0.6$ for large system size, indicating that the strong ETH typically holds for these cases~(see Supplemental Material~\cite{supplement} for a detailed analysis).
    The number of samples lies between 998 and 4994 for each data.
    Here, red~(blue) color means that the systems are likely (unlikely) to satisfy the strong ETH.
    }
\end{figure}

\paragraph{Proximity to the fully connected case.---}
To understand how the transition from the fully connected case to the short-range one occurs, we employ finite-size scaling to examine the level spacing ratio and the fractal dimension.
We first examine the level spacing ratio~\cite{oganesyan2007localization, atas2013distribution} defined by
\begin{equation}
    \tilde{r}_{\gamma} \coloneqq \min\Bqty{ \tilde{r}_{\gamma}, \frac{1}{\tilde{r}_{\gamma}} }\qc r_{\gamma} \coloneqq \frac{ E_{\gamma+1} -E_{\gamma} }{ E_{\gamma} -E_{\gamma-1} }. \label{Def:LevelSpacingRatio}
\end{equation}
The spectral average $\expval*{\tilde{r}}$ is known to be $\expval*{\tilde{r}} \simeq 0.60266$ for GUE and $\expval*{\tilde{r}} \simeq 0.38629$ for integrable systems whose level spacing distribution is Poissonian~\cite{atas2013distribution}.


Figure~\ref{fig:LevelSpacingRatioAndMultiFractalDimension}(a) shows the system-size dependence of the ensemble average $\mathbb{E}_{N}^{(\alpha)}\bqty{ \expval*{\tilde{r}} }$ for several values of $\alpha$.
For every ensemble examined $(\alpha\geq 0.25)$, it approaches the GUE value as the system size increases.
Therefore, the approximate permutation symmetry affects less for larger systems.

This result is consistent with the one for the transverse-field Ising chain with long-range interactions~\cite{russomanno2020long}.
However, for small $\alpha$, $\mathbb{E}_{N}^{(\alpha)}\bqty{ \expval*{\tilde{r}} }$ approximately lies in the middle of the GUE and Poissonian values for finite system sizes up to $N= 20$.
This fact indicates that the approximate permutation symmetry persists for small $\alpha$ in systems with a few dozens of particles.

We next evaluate the fractal dimension~\cite{backer2019multifractal} of eigenstates of a Hamiltonian $\hat{H}_{N}^{(\alpha)}[\hat{h}]$ in the eigenbasis of the corresponding fully connected Hamiltonian $\hat{H}_{N}^{(0)}[\hat{h}]$.
The fractal dimension is defined by
\begin{equation}
    D_{q}(E_{\beta}^{(\alpha)}) \coloneqq -\frac{1}{\log d_{N}} \frac{1}{q-1} \log \qty( \sum_{\gamma=1}^{d_{N}} \abs{ \braket*{ E_{\gamma}^{(0)} }{ E_{\beta}^{(\alpha)} } }^{2q} ), \label{Def:MultiFractalDimension}
\end{equation}
where $\ket*{ E_{\beta}^{(\alpha)} }$ is an eigenstate of $\hat{H}_{N}^{(\alpha)}[\hat{h}]$ with eigenenergy $E_{\beta}^{(\alpha)}$, and $\Bqty*{ \ket*{ E_{\gamma}^{(0)} } }$ is the eigenbasis of $\hat{H}_{N}^{(0)}$ to which the eigenbasis $\Bqty*{ \ket*{ E_{\gamma}^{(\alpha)} } }$ converges in the limit $\alpha\to 0$~\cite{numerics_FullyConnected}.
The fractal dimension satisfies $0\leq D_{q} \leq 1$, where the first equality holds if and only if $\abs*{ \braket*{E_{\gamma}^{(0)}}{E_{\beta}^{(\alpha)}} }^2 = 1$ for some $\gamma$, and the second equality holds if and only if $\abs*{ \braket*{E_{\gamma}^{(0)}}{E_{\beta}^{(\alpha)}} }^2 = 1/d_{N}$ for all $\gamma$
~\cite{fractalfinitesize}.

Figure~\ref{fig:LevelSpacingRatioAndMultiFractalDimension}(b) plots the ensemble average of the minimum fractal dimension $D_{2}(E_{\beta}^{(\alpha)})$ in the middle 10\% of the energy spectrum against $1/\log d_{N}$, where $d_{N}$ is the dimension of the zero-momentum even-parity sector.
For $\alpha\geq 3.0$, $D_{2}(E_{\beta}^{(\alpha)})$ approaches unity as the dimension of the Hilbert space increases, indicating that the approximate permutation symmetry disappears for sufficiently large system size.
The data for $\alpha=1.0$ also tends to approach unity, albeit slowly.

Although the fractal dimension for $\alpha= 0.5$ slightly increases for $1/\log d_N\geq 0.09~(N\leq 20)$, its slope is not large enough to determine whether it approaches unity or converges to a smaller value.
For ensembles with $\alpha= 0.25$, $D_{2}(E_{\beta}^{(\alpha)})$ does not increase within computationally accessible system size ($N\leq 20$), suggesting that it remains small for larger system size.
Thus, eigenstates of Hamiltonians with $\alpha\lesssim 0.5$ retain some resemblance to those of the fully connected Hamiltonian even for large system size.
Since the eigenstates of a fully connected Hamiltonian typically violate the strong ETH, the eigenstate expectation values for $\alpha\lesssim 0.5$ are expected to deviate from the microcanonical average even for relatively large system sizes due to the proximity to the fully connected Hamiltonian.

\begin{figure}[tb]
    \centering
    \includegraphics[width=\linewidth]{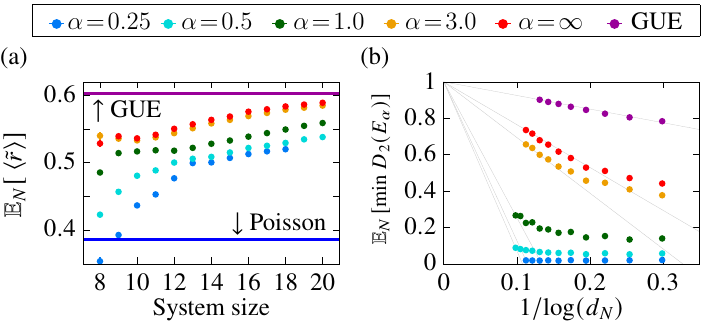}
    \caption{ \label{fig:LevelSpacingRatioAndMultiFractalDimension}
        (a) Ensemble average of the mean level spacing ratio $\expval*{\tilde{r}}$ defined in Eq.~\eqref{Def:LevelSpacingRatio}, where the average $
        \expval*{\cdots}$ is taken over the middle 10\% of the spectrum.
        It approaches the GUE value for all $\alpha$ with increasing system size.
        (b) Ensemble average of the minimum fractal dimension with $q=2$ for energy eigenstates of $\hat{H}_{N}^{(\alpha)}$ with respect to the eigenbasis of the fully connected Hamiltonian $\hat{H}_{N}^{(0)}$.
        The minimum is taken over the middle 10\% of the spectrum.
        It approaches unity for $\alpha\geq 1.0$ with increasing $d_{N}$.
        Whether the data for $\alpha=0.5$ approaches unity or converges to a smaller value is unclear.
        For $\alpha=0.25$, the minimum fractal dimension shows no increase.
        Each gray line connects the point $(0,1)$ and the data point with the largest $d_{N}$.
        The number of samples lies between 996 and 4994.
        Most error bars are smaller than the dot size.
    }
\end{figure}

\paragraph{Range of validity of Srednicki's ansatz.---}
We test the validity of the first part (Eq.~\eqref{eq:SrednickiAnsatz}) of Srednicki's ansatz~(see Supplemental Material~\cite{supplement} for the second).
By applying Boltzmann's formula $S(E) \sim \log d_{E,\delta E}$ with $d_{E,\delta E} \coloneqq \dim \mathcal{H}_{E,\delta E}$ to Eq.~\eqref{eq:SrednickiAnsatz}, we obtain $\mathcal{S}\bqty*{ \delta O_{\gamma\gamma} } \simeq (\sqrt{d_{E_{\gamma},\delta E}})^{-1} f(E_{\gamma})$~\cite{srednickiscaling}.
We test Eq.~\eqref{eq:SrednickiAnsatz} for our ensembles by investigating the $d_{E,\delta E}$-dependence of the estimator $\hat{\mathcal{S}}^{E}_{\delta E}$ of $\eval{\mathcal{S}\bqty*{ \delta O_{\gamma\gamma} }}_{E_{\gamma} \simeq E} = \eval{ \sqrt{\mathcal{E}\bqty*{ (\delta O_{\gamma\gamma})^2 } }}_{E_{\gamma} \simeq E}$ defined by
\begin{equation}
    \hat{\mathcal{S}}^{E}_{\delta E} \coloneqq \sqrt{ \frac{1}{d_{E,\delta E}} \sum_{ \ket*{E_{\gamma}} \in \mathcal{H}_{E,\delta E}} (\delta O_{\gamma\gamma})^2 }.
\end{equation}

For each sample $(\hat{h}, \hat{o}) \in \mathcal{L}(\mathcal{H}_{\mathrm{loc}}^{\otimes 2}) \times \mathcal{L}(\mathcal{H}_{\mathrm{loc}}^{\otimes 2})$, we construct a Hamiltonian $\hat{H}_{N}^{(\alpha)}[\hat{h}]$ and an observable $\hat{O}_{N}^{(\infty)}[\hat{o}]$ as in Eq.~\eqref{Eq:Hamiltonian} for various $N$ and
fit the numerically obtained $\hat{\mathcal{S}}^{E}_{\delta E}$ with a function $C (\sqrt{ d_{E,\delta E} })^{-a}$ by appropriately choosing parameters $C$ and $a$ (note that $d_{E,\delta E}$ depends on $N$).
The validity of this fitting is tested by comparing its mean squared residual with that of the fitting with a function $C' (\log d_{E,\delta E})^{-a'}$, which applies to the integrable case.
The probability distributions of $a$ for different $\alpha$ are shown in Fig.~\ref{fig:ExponentHist_Pow}.

If Srednicki's ansatz holds typically, we have $a\sim 1$ with high probability; therefore, the probability distribution of $a$ should peak around unity.
To estimate finite-size effects, we restrict the available system size for the fitting of $\hat{\mathcal{S}}^{E}_{\delta E}$ with $C (\sqrt{ d_{E,\delta E} })^{-a}$ to $N_{\max}$ and vary $N_{\max}$.
For $\alpha=3.0$, the probability density tends to peak around $a=1$ and decreases for small $a$ as $N_{\max}$ increases.
We find a similar tendency for $\alpha\gtrsim 1.2$~(see Supplemental Material~\cite{supplement}).
Therefore, the first part of Srednicki's ansatz typically holds in the thermodynamic limit for $\alpha\gtrsim 1.2$.

However, the finite-size-scaling behavior for $\alpha\leq 1$ shows no tendency for the distribution to peak around unity, indicating the breakdown of Srednicki's ansatz at least for relatively large system sizes.
For small $\alpha\, (\lesssim 0.5)$, $C' (\log d_{E,\delta E})^{-a'}$ fits the data as well as $C (\sqrt{ d_{E,\delta E} })^{-a}$.
This fact indicates that the peaks of the distributions for $\alpha=0.5$ and $\alpha\simeq 0$ in Fig.~\ref{fig:ExponentHist_Pow} are artifacts of an improper fitting to $C (\sqrt{ d_{E,\delta E} })^{-a}$, which always yields a positive value of $a$ whenever $\hat{\mathcal{S}}^{E}_{\delta E}$ decreases with increasing $d_{E,\delta E}$.

\begin{figure}[tb]
    \centering
    \includegraphics[width=\linewidth]{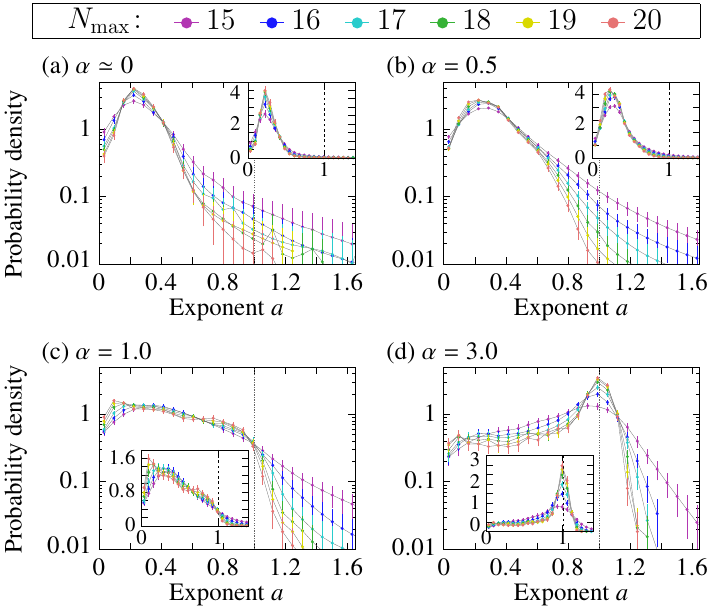} 
    \caption{ \label{fig:ExponentHist_Pow}
        Distribution of the exponent $a$ in the fitting $\hat{\mathcal{S}}^{E}_{\delta E} \propto (\sqrt{d_{E,\delta E}})^{-a}$.
        The inset shows the same data in linear scale.
        The existence of a clear peak around $a=1$ shows that Srednicki's ansatz holds for $\alpha=3.0$.
        No peak around $a=1$ can be found for $\alpha\leq 1.0$ even for the largest available system size, indicating the breakdown of Srednicki's ansatz at least for relatively large system sizes.
        The number of samples lies between 1000 and 5000.
    }
\end{figure}

Srednicki's ansatz is based on the observation that the relationship of a quantum many-body Hamiltonian to a physical observable resembles that between two Gaussian random matrices~
\cite{beugeling2014finite}.
To check this for long-range interactions, we examine the system-size dependence of the fractal dimension~\eqref{Def:MultiFractalDimension} of eigenstates of $\hat{H}_{N}^{(\alpha)}$ with $\alpha\in[0,3]$ in the eigenbasis of a local operator $\hat{O}_{N}^{(\infty)}$, i.e., we replace $\Bqty*{ \ket*{ E_{\gamma}^{(0)} } }$ in Eq.~\eqref{Def:MultiFractalDimension} with the eigenbasis of $\hat{O}_{N}^{(\infty)}$.
The results are shown in Fig.~\ref{fig:MFD_vsOp}.
For $\alpha\geq 3$, where the typicality of both the strong ETH and Srednicki's ansatz has been established in Ref.~\cite{sugimoto2021test} and Fig.~\ref{fig:ExponentHist_Pow}, we find that the fractal dimension approaches unity as the system size increases.
However, the fractal dimension increases rather slowly for $\alpha=1.0$ and decreases for $\alpha\leq 0.5$.
This result implies a strong correlation between eigenstates of a Hamiltonian and those of a local observable when the interactions are long-ranged,
invalidating the application of the conventional random matrix theory for $\alpha\lesssim 1.0$.

\begin{figure}[tb]
    \centering
    \includegraphics[width=\linewidth]{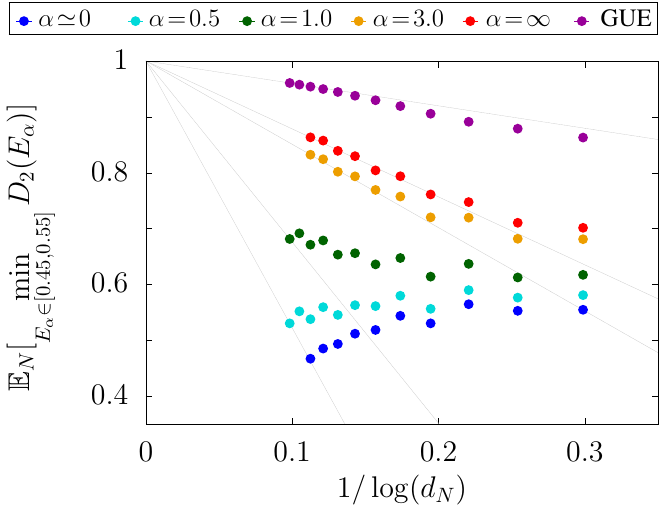}
    \caption{\label{fig:MFD_vsOp}
        Ensemble average of the minimum fractal dimension~\eqref{Def:MultiFractalDimension} of energy eigenstates with respect to the eigenbasis of a local observable $\hat{O}$.
        While it approaches unity for $\alpha\geq 3.0$, it increases rather slowly for $\alpha=1.0$ and decreases for $\alpha\leq 0.5$, indicating a strong correlation between long-ranged Hamiltonians and local observables.
        The number of samples ranges from 932 to 2000 for all data points.
    }
\end{figure}

\paragraph{Conclusion.---}
We have found that the strong ETH typically holds for one-dimensional systems with two-body long-range interactions~$1/r^{\alpha}$ at least for $\alpha\gtrsim 0.6$, which include important cases of Coulomb ($\alpha=1$), monopole-dipole ($\alpha=2$), and dipole-dipole~($\alpha=3$) interactions.
We have also shown that generic two-body long-range interactions make $\mathbb{E}_{N}^{(\alpha)}[\Delta_{\infty}]$ significantly larger than that for short-range interacting systems.
Indeed, we cannot decide whether or not the strong ETH typically holds for $\alpha\leq 0.5$ within the computationally available system sizes~($N\leq 20$).
These results are directly relevant for understanding thermalization dynamics of finite-size systems realizable in actual experiments.
We find that Srednicki's ansatz typically holds for $\alpha\gtrsim 1.2$ but typically breaks down for $\alpha\lesssim 1.0$ for computationally tractable system size.
Our results reveal a region $(0.5\lesssim \alpha \lesssim 1.0)$ where the strong ETH typically holds, but Srednicki's ansatz typically breaks down.

Thus, not only the experimentally investigated long-range Ising interaction~\cite{neyenhuis2017observation} but also \textit{generic} long-range interactions impede thermalization.
We have studied the dynamics of long-range interacting systems from simple initial states with energy expectation values in the middle 20\% of the spectrum and found that the equilibrium expectation value of a short-range observable typically deviates more from the microcanonical average for smaller $\alpha$~\cite{supplement}.

The critical value $\alpha_{c}=1.0$ below which Srednicki's ansatz typically breaks down for one-dimensional systems is precisely the value below which the additivity of a physical quantity is lost.
Given the importance of additivity in thermodynamics, we expect that the strong ETH and Srednicki's ansatz typically hold at least when the range of interactions is shorter than $1/r^{d}$ for $d$-dimensional systems.
It remains a challenge to clarify the relationship between the additivity and the strong ETH, and how the critical value of $\alpha$ changes for higher dimensions.

\begin{acknowledgments}
We are very grateful to Synge Todo and Tilman Hartwig for their help in our numerical calculation.
We also thank Liu Ziyin for helpful discussions in the statistical analysis.
This work was supported by KAKENHI Grant Numbers JP18H01145 from the Japan Society for the Promotion of Science (JSPS). 
S.~S. was supported by Forefront Physics and Mathematics Program to Drive Transformation (FoPM), a World-leading Innovative Graduate Study (WINGS) Program, the University of Tokyo.
\end{acknowledgments}

\bibliography{reference}

\begin{thebibliography}{91}
\bibitem [{bas()}]{basischoice}%
  \BibitemOpen
  \href@noop {} {}\bibinfo {note} {{Because the probabilistic measure $P$ we
  introduce on the ensemble $\mathcal{G}^{(\alpha)}$ does not depend on the
  choice of an orthonormal operator basis $\Bqty*{ \hat{\sigma}^{(p)} }$, we
  can arbitrarily choose it }}\BibitemShut {NoStop}%
\bibitem [{tim()}]{timeReversalSymmetry}%
  \BibitemOpen
  \href@noop {} {}\bibinfo {note} {{On the other hand, it is not invariant
  under time-reversal operation because it contains terms like
  $\hat{\sigma}^{x}_{j}\hat{\sigma}^{y}_{k}$ }}\BibitemShut {NoStop}%
\bibitem [{mea()}]{measureinvariance}%
  \BibitemOpen
  \href@noop {} {}\bibinfo {note} {{This measure is invariant under any change
  of the local orthonormal operator basis $\Bqty*{ \hat{\sigma}^{(p)}
  }_{p=0}^{d_{L}^2-1}$ }}\BibitemShut {NoStop}%
\bibitem [{num()}]{numerics_FullyConnected}%
  \BibitemOpen
  \href@noop {} {}\bibinfo {note} {{In the numerical calculation of $\Bqty*{
  \ket*{ E_{\beta}^{(0)} } }_{\beta}$, we take $\alpha=0.0001$ to break the
  degeneracies. This value is small enough compared with the system size of the
  simulation and essentially describes the physics of the fully connected
  case~($\alpha=0$) }}\BibitemShut {NoStop}%
\end{thebibliography}

\clearpage\clearpage
\makeatletter
   	\c@secnumdepth=4
    \def\@pointsize{11}
	\expandafter\@process@pointsize\expandafter{\@pointsize@default}%
	\appdef\setup@hook{\normalsize}%
	\setup@hook
\makeatother

\setcounter{equation}{0}
\setcounter{figure}{0}
\setcounter{section}{0}
\setcounter{table}{0}
\renewcommand{\theequation}{S\arabic{equation}}
\renewcommand{\thefigure}{S\arabic{figure}}
\renewcommand{\theHequation}{\theequation}
\renewcommand{\theHfigure}{\thefigure}

\title{
    Supplemental Material: \protect\\
    Eigenstate Thermalization in Long-Range Interacting Systems
}
\date{\today}
\maketitle
\onecolumngrid

\section{Translation invariance and parity invariance of the model}
In this section, we prove that the operators introduced in Eq.~\eqref{Eq:Hamiltonian} in the main text, which can be rewritten as
\begin{gather}
    \hat{H}_{N}^{(\alpha)}[\hat{h}] \coloneqq \sum_{p,q=1}^{ d_{L}^2-1 } h_{pq}\, \hat{\Lambda}_{pq,N}^{(\alpha)}\qc \nonumber \\ 
    \hat{\Lambda}_{pq,N}^{(\alpha)} \coloneqq \sum_{ \substack{j,k=0 \\ j\neq k} }^{N} \frac{ \hat{\sigma}^{(p)}_{j} \hat{\sigma}^{(q)}_{k} }{(r_{jk})^{\alpha}}\qc 
    r_{jk} \coloneqq \min\Bqty{ \abs*{j-k}, N-\abs*{j-k} }, \label{Eq:SMBasisOperator}
\end{gather}
are invariant under the translation $\hat{T}_{N}\hat{\sigma}^{(p)}_{j}\hat{T}_{N}^{\dagger} = \hat{\sigma}^{(p)}_{j+1}$ and the parity transformation $\hat{P}_{N}\hat{\sigma}^{(p)}_{j}\hat{P}_{N}^{\dagger} = \hat{\sigma}^{(p)}_{N+1-j}$.
The translation and parity invariance of $\hat{H}_{N}^{(\alpha)}[\hat{h}]$ originates from those of $\hat{\Lambda}_{pq,N}^{(\alpha)}$ as shown below.

The transformation invariance of $\hat{\Lambda}_{pq,N}^{(\alpha)}$ can be confirmed by rewriting the sum $\sum_{ j\neq k }$ as $\sum_{ j } \sum_{ \Delta\neq 0}$ where $\Delta \coloneqq k-j$.
Indeed, we obtain
\begin{align}
    \hat{\Lambda}_{pq,N}^{(\alpha)}
    &= \sum_{j=1}^{N} \sum_{\Delta \neq 0} \frac{ \hat{\sigma}^{(p)}_{j} \hat{\sigma}^{(q)}_{j+\Delta} }{ (r_{0,\Delta})^{\alpha} }
    = \sum_{j=1}^{N} (\hat{T}_{N})^{j-1} \qty( \sum_{\Delta \neq 0} \frac{ \hat{\sigma}^{(p)}_{1} \hat{\sigma}^{(q)}_{1+\Delta} }{ (r_{0,\Delta})^{\alpha} } ) (\hat{T}_{N}^{\dagger})^{j-1}.
\end{align}
Since the sum $\sum_{ j=1 }^{N}$ runs over all the sites, the rightmost-hand side is translation invariant, and so does the leftmost one.

The parity invariance of $\hat{\Lambda}_{pq,N}^{(\alpha)}$ can be confirmed by changing the variables $(j,k)$ to $(N+1-j,N+1-k)$, obtaining
\begin{align}
    \hat{P}_{N}\hat{\Lambda}_{pq,N}^{(\alpha)}\hat{P}_{N}^{\dagger}
    &= \sum_{  \substack{j,k=0 \\ j\neq k} }^{N} \frac{ \hat{\sigma}^{(p)}_{N+1-j} \hat{\sigma}^{(q)}_{N+1-k} }{ (r_{j,k})^{\alpha} }
    = \sum_{  \substack{j,k=0 \\ j\neq k} }^{N} \frac{ \hat{\sigma}^{(p)}_{j} \hat{\sigma}^{(q)}_{k} }{ (r_{N+1-j,N+1-k})^{\alpha} }
    = \sum_{  \substack{j,k=0 \\ j\neq k} }^{N} \frac{ \hat{\sigma}^{(p)}_{j} \hat{\sigma}^{(q)}_{k} }{ (r_{j,k})^{\alpha} } = \hat{\Lambda}_{pq,N}^{(\alpha)},
\end{align}
where we have changed the variables in the second equality, and we have used $r_{N+1-j,N+1-k} = r_{j,k}$ in the third equality.

\section{Supplement to the section ``Finite-size scaling of the strong ETH measure" in the main text}
\subsection{Odd-even staggering of $\mathbb{E}_{N}[\Delta_{\infty}]$}
\begin{figure}[tbh]
    \centering
    \includegraphics[width=\linewidth]{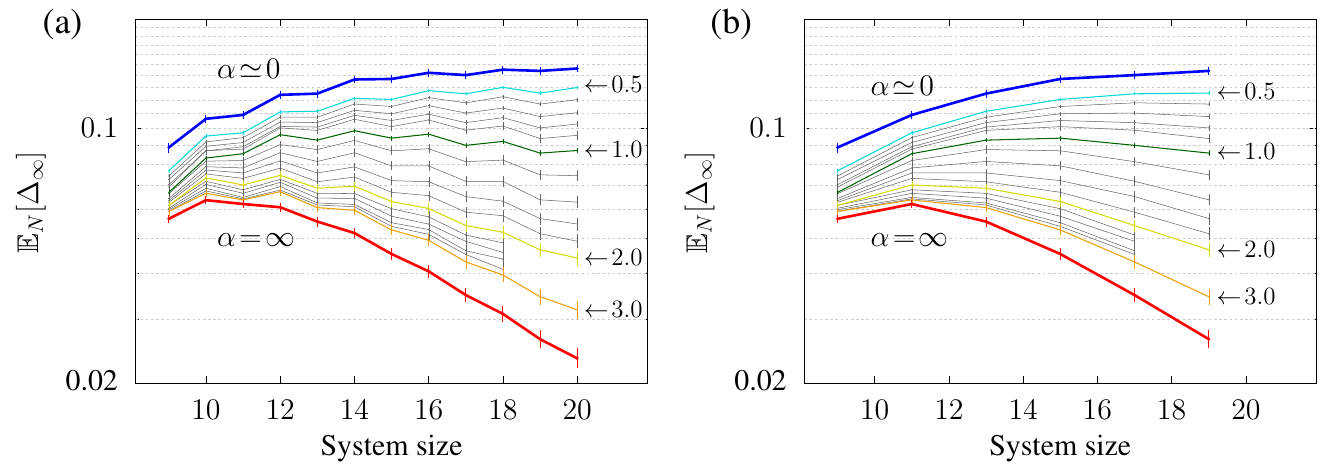}
    \caption{\label{fig:SMOdd_Even_Staggering}
        Odd-even staggering of the ensemble average $\mathbb{E}_{N}[\Delta_{\infty}]$ as a function of the system size $N$.
        (a) Data for both odd and even $N$.
        The ensemble average $\mathbb{E}_N[\Delta_\infty]$ exhibits staggering behavior.
        (b) The data for only odd $N$, where the average $\mathbb{E}_{N}[\Delta_{\infty}]$ varies smoothly.
        The gray curves between $\alpha = 0.5$ and $1.0$ show the data for $\alpha=0.6,0.7,0.8,0.9$, and those between $\alpha = 1.0$ and $3.0$ are for $\alpha=1.2,1.4,\dots,2.8$.
    }
\end{figure}

The ensemble average $\mathbb{E}_{N}[\Delta_{\infty}]$ of the measure $\Delta_{\infty}$ shows odd-even staggering behavior possibly because of the finite-size effect and the even-odd parity effect as a function of the system size $N$ as shown in Fig.~\ref{fig:SMOdd_Even_Staggering}(a).
If we focus only on odd or even $N$, a smooth behavior is observed as in Fig.~\ref{fig:SMOdd_Even_Staggering}(b) for odd $N$ and Fig.~1 for even $N$ in the main text.

\clearpage
\subsection{Fitting the numerical data with the function in Eq.~(5) in the main text}
\begin{figure}[hb]
    \centering
    \includegraphics[width=\linewidth]{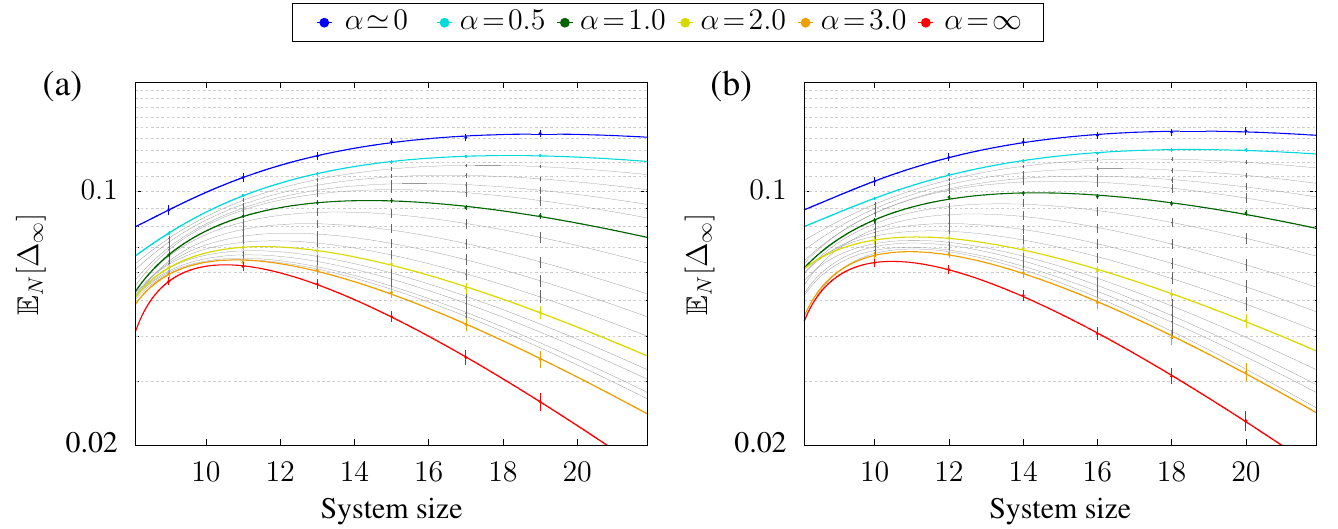}
    \caption{\label{fig:SMFitting}
        Ensemble average of the measure of the strong ETH $\Delta_{\infty}$ defined in Eq.~(1) in the main text.
        The color code is the same as in Fig.~1(a) in the main text.
        Each curve is the asymptotic formula~\eqref{Eq:SMAsymptoticFormula} fitted to the numerical data by adjusting $C$, $N_{m}$ and $N_{0}$.
        The formula~\eqref{Eq:SMAsymptoticFormula} fits quite well to the numerical data for all the values of $\alpha\geq 0$ \textit{including} $\alpha=0$, whereas the strong ETH typically breaks down for $\alpha=0$ because of the permutation symmetry of any two sites.
    }
\end{figure}

For Gaussian random matrix ensembles, where the locality and few-bodiness of realistic operators are completely disregarded, the asymptotic dependence of $\mathbb{E}_{N}\bqty*{ \Delta_{\infty} }$ is obtained as
\begin{equation}
    \mathbb{E}_{N}^{(\mathrm{RMT})}\bqty*{ \Delta_{\infty} }
\simeq C Ne^{ -N/N_{m} } \sqrt{ 1-\frac{N_{m}}{2} \frac{\log N}{N} -\frac{N_{0}}{N} }, \label{Eq:SMAsymptoticFormula}
\end{equation}
where $C$, $N_{m}$, and $N_{0}$ are constants~\citeSM{sugimoto2021testSM}.
This function fits quite well to numerically obtained $\mathbb{E}_{N}^{(\mathrm{RMT})}\bqty*{ \Delta_{\infty} }$ for all the values of $\alpha$ \textit{including} $\alpha = 0$ as shown in Fig.~\ref{fig:SMFitting}.

However, as argued in the main text, for $\alpha = 0$, the strong ETH typically breaks down because of the permutation symmetry of any two sites.
Therefore, whether or not Eq.~\eqref{Eq:SMAsymptoticFormula} fits well to the numerical data does not faithfully reflect whether or not $\mathbb{E}_{N}[\Delta_{\infty}]$ vanishes in the thermodynamics limit, and we need to check whether the numerically obtained $\mathbb{E}_{N}[\Delta_{\infty}]$ actually decreases for large $N$ as done by using the bootstrap method.

\clearpage
\subsection{Estimation of the statistical error}
Suppose that we sample $M$ elements $\Bqty{ x_{j} \coloneqq (\hat{h}_{j}, \hat{o}_{j}) }_{j=1}^{M}$ from $\mathcal{L}(\mathcal{H}_{\mathrm{loc}}^{\otimes 2}) \times \mathcal{L}(\mathcal{H}_{\mathrm{loc}}^{\otimes 2})$.
For each sample $x_{j} = (\hat{h}_{j}, \hat{o}_{j})$, we construct a Hamiltonian $\hat{H}_{N}^{(\alpha)}[\hat{h}]$ and an observable $\hat{O}_{N}^{(\infty)}[\hat{o}]$ for various $N$ as in Eq.~(3) in the main text and obtain a sequence $\Bqty{ (\Delta_{\infty})_{N}^{(j)} }_{N}$ of the measure of the strong ETH.
From these samples, we estimate the ensemble average $\mathbb{E}_{N}[\Delta_{\infty}]$ by
\begin{equation}
    \hat{\mathbb{E}}_{N}[\Delta_{\infty}] = \frac{1}{M} \sum_{j=1}^{M} (\Delta_{\infty})_{N}^{(j)}. \label{Eq:SMEstimator}
\end{equation}
This estimator depends on the realization of the samples $\Bqty{ x_{j} }_{j=1}^{M}$, and we have to estimate the statistical error contained in the estimation of $\mathbb{E}_{N}[\Delta_{\infty}]$.
Therefore, we quantify the level of confidence about our results with the bootstrap method~\citeSM{efron1994introduction} explained below.

We randomly choose $M$ samples from $\Bqty{ x_{j} }_{j=1}^{M}$ allowing repetitions, and denote them as $\Bqty*{ \tilde{x}_{j}^{(\alpha)} }_{j=1}^{M}\ (\alpha=1,\dots,B)$, where $B \ (=10000)$ is the number of the bootstrap iterations.
Then, we calculate the estimator~\eqref{Eq:SMEstimator} from $\Bqty*{ \tilde{x}_{j}^{(\alpha)} }_{j=1}^{M}$, obtaining $\hat{\mathbb{E}}_{N_\mathrm{min}}[\Delta_{\infty}]^{(\alpha)}, \cdots ,
\hat{\mathbb{E}}_{N_\mathrm{max}}[\Delta_{\infty}]^{(\alpha)}$.
By repeating this procedure $B$ times, we obtain the joint distribution of the estimators $\hat{\mathbb{E}}_{N_{\min}}[\Delta_{\infty}], \cdots, \hat{\mathbb{E}}_{N_{\max}}[\Delta_{\infty}]$. 
We then calculate the probability $P_{\mathrm{dec}}$ of obtaining a sequence such that $\hat{\mathbb{E}}_{N_{\min}}[\Delta_{\infty}] > \hat{\mathbb{E}}_{N_{\min}+2}[\Delta_{\infty}] > \dots > \hat{\mathbb{E}}_{N_{\max}}[\Delta_{\infty}]$.
We use this probability as the level of confidence about the fact that the ensemble average $\mathbb{E}_{N}[\Delta_{\infty}]$ decreases with increasing $N$.

\begin{figure}[bth]
    \centering
    \includegraphics[width=\linewidth]{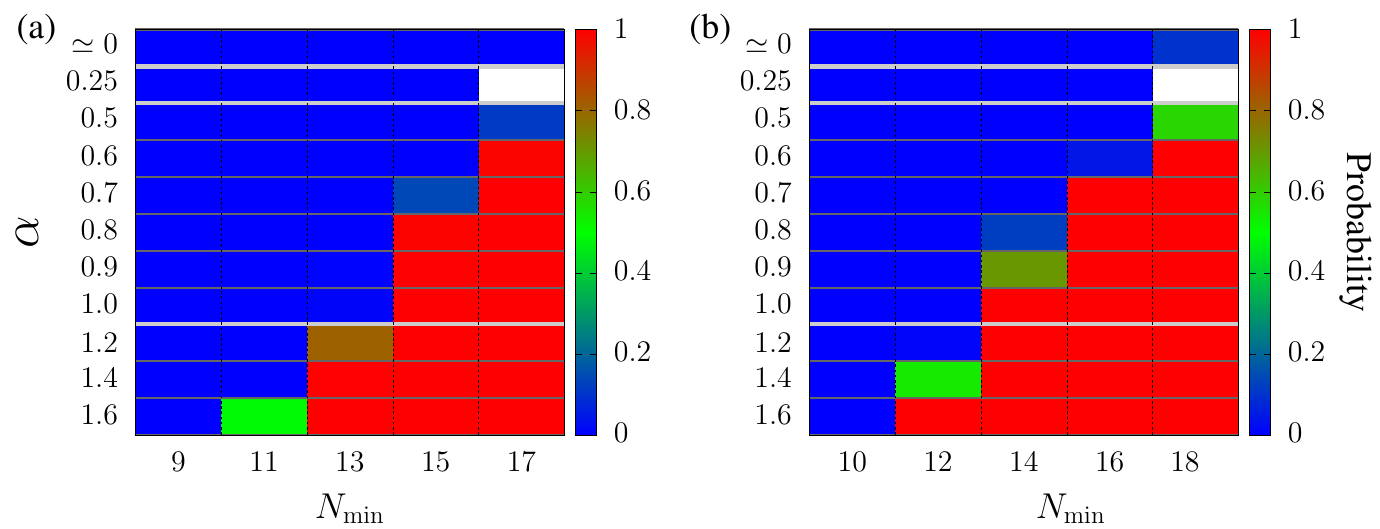}
    \caption{\label{fig:SMBootstrapProbability}
    Probability of a sequence $\Bqty{ \hat{\mathbb{E}}_{N}[\Delta_{\infty}] }_{N=N_{\min}}^{N_{\max}}$ being obtained such that $\hat{\mathbb{E}}_{N_{\min}}[\Delta_{\infty}] > \hat{\mathbb{E}}_{N_{\min}+2}[\Delta_{\infty}] > \dots > \hat{\mathbb{E}}_{N_{\max}}[\Delta_{\infty}]$ as represented by the color of each cell.
    White cells indicate where no data is available.
    The number of samples $M$ used to calculate each estimator $\hat{\mathbb{E}}_{N}[\Delta_{\infty}]$ ranges from 997 to 3307 depending on $\alpha$, and the number of bootstrap iterations $B$ is 10000.
    We set $N_{\max} = 20 \ (19)$ for $\alpha\neq0.25$ and $N_{\max} = 18 \ (17)$ for $\alpha=0.25$ for even (odd) $N$.
    The panel (a) shows the data for odd $N$, and the panel (b) shows the data for even $N$.
    The average $\mathbb{E}_{N}[\Delta_{\infty}]$ starts decreasing for $\alpha\geq 0.6$ for the system size used in our calculation, which implies the typicality of the strong ETH.
    For $\alpha=0.5$, we have $P_{\mathrm{dec}} = 0.1140$ for odd $N$ with $N_{\min}=17$ and $P_{\mathrm{dec}} = 0.5803$ for even $N$ with $N_{\min}=18$.
    This result implies that we cannot decide whether $\mathbb{E}_{N}[\Delta_{\infty}]$ decreases or increases for $N\geq 18$ within the available system size.
    }
\end{figure}

The result is shown in Fig.~\ref{fig:SMBootstrapProbability} for various $N_{\min}$ with $N_{\max}$ being the maximum system size calculated for each $\alpha$, i.e., $N_{\max} = 20 \ (19)$ for $\alpha\neq0.25,1.4,1.6$ and $N_{\max} = 18 \ (17)$ for $\alpha=0.25,1.4,1.6$ for even (odd) $N$.
We observe that the probability of $\Bqty{ \hat{\mathbb{E}}_{N}[\Delta_{\infty}] }_{N=N_{\min}}^{N_{\max}}$ monotonically decreasing with increasing $N$ is (almost) unity for $\alpha\geq 0.6$.
Specifically, we obtain $P_{\mathrm{dec}} = 0.9999$ for $\alpha=0.6$ with $N_{\min}=18$, and $P_{\mathrm{dec}} = 1.0$ for $\alpha\geq 0.7$.
For all the other red cells, we have $P_{\mathrm{dec}}\geq 0.94$.
This result implies that we can safely judge from the data that $\mathbb{E}_{N}[\Delta_{\infty}]$ decreases for $\alpha\geq 0.6$.
On the other hand, for $\alpha=0.5$, we have $P_{\mathrm{dec}}\geq 0.6683$ for $N_{\min}=17$ and $P_{\mathrm{dec}}\geq 0.4467$ for $N_{\min}=18$.
These results imply that we cannot determine from the available data whether or not $\mathbb{E}_{N}[\Delta_{\infty}]$ decreases.

\begin{figure}[tbh]
    \centering
    \includegraphics[width=\linewidth]{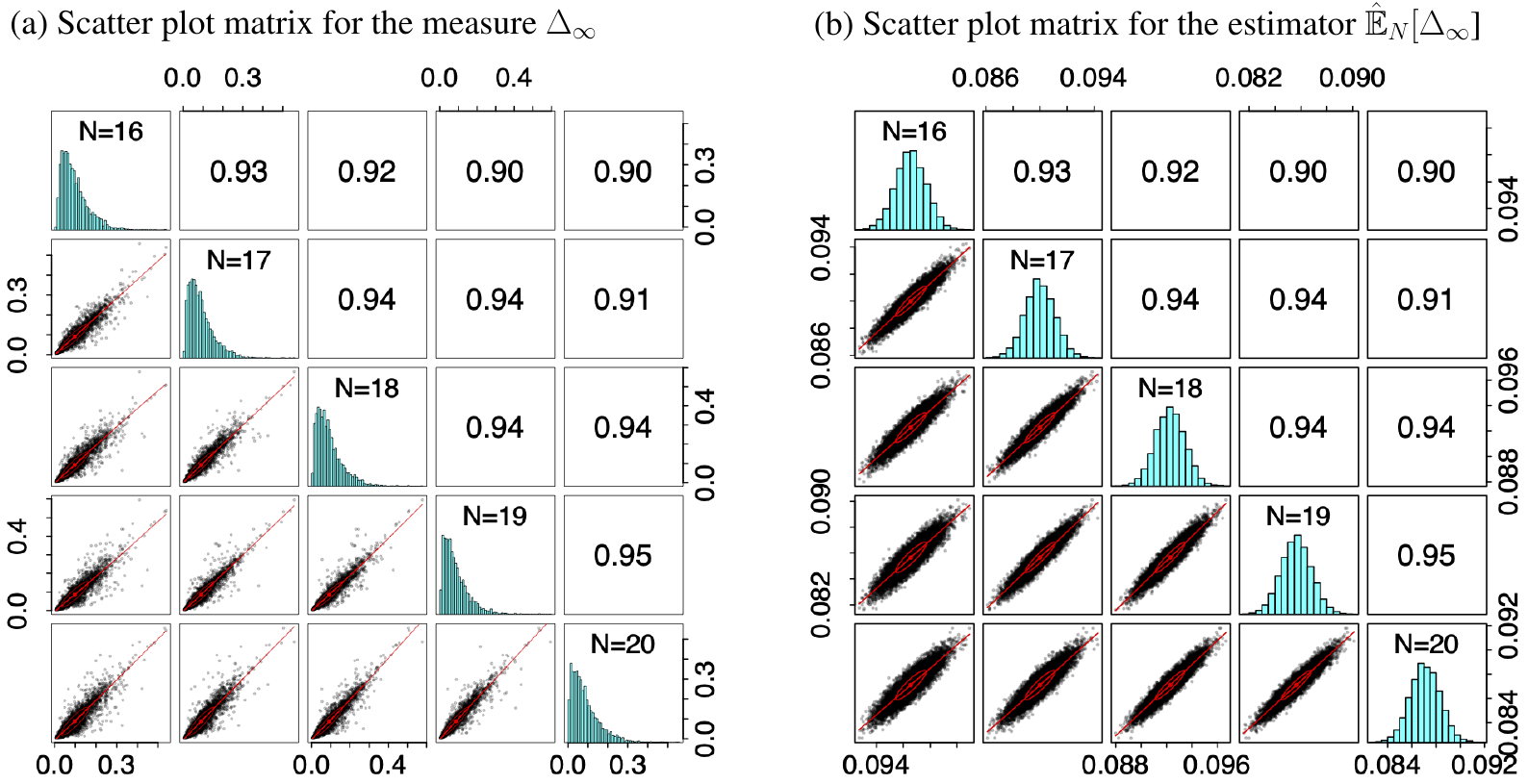}
    \caption{\label{fig:SMScatterPlot}
    Scatter plot matrix for (a) the measure $\Delta_{\infty}$ in Eq.~(1) in the main text calculated from each sample, and that for (b) the estimator $\hat{\mathbb{E}}_{N}[\Delta_{\infty}]$ in Eq.~\eqref{Eq:SMEstimator} obtained for each bootstrap iteration for $\alpha=1.0$.~(Similar results are obtained for other $\alpha$.)
    The system size $N$ for each row and column is indicated in the diagonal panels.
    For both (a) and (b), the lower off-diagonals show the scatter plots, the diagonals show the histograms, and the upper off-diagonals show the Pearson correlation.
    Strong correlations of $\Delta_{\infty}$ and $\hat{\mathbb{E}}_{N}[\Delta_{\infty}]$ between different $N$ strengthen our conclusion that $\hat{\mathbb{E}}_{N}^{(\alpha)}[\Delta_{\infty}]$ for $\alpha\geq 0.6$ vanishes in the thermodynamics limit which is drawn from a finite number of samples.
    }
\end{figure}

Finally, we note that there is a strong correlation between $(\Delta_{\infty})_{N}$ for different $N$ obtained from the same sample $x_{j} = (\hat{h}_{j}, \hat{o}_{j})$ as shown in Fig.~\ref{fig:SMScatterPlot}. 
This correlation enables us to almost certainly decide if $\mathbb{E}_{N}[\Delta_{\infty}]$ decreases for $\alpha\geq 0.6$, although the width of the $80\%$ confidence interval of $\mathbb{E}_{N}[\Delta_{\infty}]$ for each $N$ is not negligible for $\alpha\leq0.8$ as shown in Fig.~1 in the main text.
Indeed, if the estimator $\hat{\mathbb{E}}_{N}[\Delta_{\infty}]$ for different $N$ were independent of one another, the probability of obtaining a monotonically decreasing sequence $\Bqty{ \hat{\mathbb{E}}_{N}[\Delta_{\infty}] }_{N=N_{\min}(\leq 18)}^{N_{\max}(=20)}$ for $\alpha\leq 0.8$ would become smaller than $80\%$, as suggested by the overlap of the $80\%$ confidence intervals for different $N$ in Fig.~1 in the main text.
Figure~\ref{fig:SMScatterPlot} also shows that the distribution of $\Delta_{\infty}$ is not Gaussian, while the number of samples $M (\geq 1000)$ in our calculation is sufficiently large so that the distribution of $\hat{\mathbb{E}}_{N}[\Delta_{\infty}]$ becomes Gaussian.

\vspace{3cm}

\clearpage
\subsection{Relaxation dynamics from relatively simple initial states}
In this section, we present the absence of thermalization from relatively simple initial states in our setting.
Since we restrict ourselves to the zero-momentum and even-parity sector, we adopt the following states as our initial states
\begin{align}
    \ket*{\vec{\sigma};0,+} &\coloneqq 
    \begin{cases}
    \begin{aligned}
        &\ket*{\vec{\sigma};0} & ( \hat{P}_{N}\ket*{\vec{\sigma};0} = \ket*{\vec{\sigma};0} ); \\[1ex]
        &\frac{ \ket*{\vec{\sigma};0} +\hat{P}_{N} \ket*{\vec{\sigma};0} }{ \sqrt{2} } & ( \hat{P}_{N}\ket*{\vec{\sigma};0} \neq \ket*{\vec{\sigma};0} ),
    \end{aligned}
    \end{cases} \nonumber \\
    \ket*{\vec{\sigma};0} &\coloneqq \frac{1}{ \sqrt{p(\vec{\sigma})} } \sum_{j=1}^{p(\vec{\sigma})} \hat{T}_{N}^{j} \ket*{\vec{\sigma}},
\end{align}
where $\hat{P}_{N}$ and $\hat{T}_{N}$ are the parity and the translation operators, respectively, $\ket*{\vec{\sigma}} \coloneqq \ket*{\sigma_{1}}\otimes\dots\otimes\ket*{\sigma_{N}}$ with $\hat{\sigma}^{(3)}_{j} \ket*{\sigma_{j}} = \sigma_{j} \ket*{\sigma_{j}}\ (\sigma_{j} = 0,1)$, and $p(\vec{\sigma})$ is the smallest positive integer such that $\hat{T}_{N}^{p(\vec{\sigma})} \ket*{\vec{\sigma}} = \ket*{\vec{\sigma}}$.

We seek for a state $\ket{\vec{\sigma};0;+}$ where the energy (including the standard deviation) lies within the middle 20\% of the spectrum\footnote{
In our work, we test the strong ETH in the middle 10\% of the energy spectrum.
However, we have found that states $\ket*{\vec{\sigma}; 0,+}$ whose energy lies within the middle 10\% of the spectrum (in a similar sense as in Eq.~\eqref{Eq:SMEnergy}) do not exist for most of the samples. 
Therefore, we have instead adopted the condition~\eqref{Eq:SMEnergy}.
};
\begin{gather}
    E_{ \vec{\sigma} } \pm \delta E_{ \vec{\sigma} } \in [0.4\eta_{H},0.6\eta_{H}], \nonumber \\
    E_{ \vec{\sigma} } \coloneqq \expval*{ \hat{H} }{ \vec{\sigma}; 0,+ }\qc
    \delta E_{ \vec{\sigma} } \coloneqq \sqrt{ \expval*{ \hat{H}^2 }{ \vec{\sigma}; 0,+ } - \expval*{ \hat{H} }{ \vec{\sigma}; 0,+ }^2 }. \label{Eq:SMEnergy}
\end{gather}
Here, the state $\ket*{ \vec{\sigma}; 0,+ }$ satisfying the condition~\eqref{Eq:SMEnergy} does not always exist.
In Table~\ref{tab:SMNumberOfSamples}, we list the total number of samples calculated in our numerical simulation and the number of samples for which the state $\ket*{ \vec{\sigma}; 0,+ }$ satisfying the condition~\eqref{Eq:SMEnergy} exists.

\begin{figure}[hbt]
    \centering
    \includegraphics[width=0.48\linewidth]{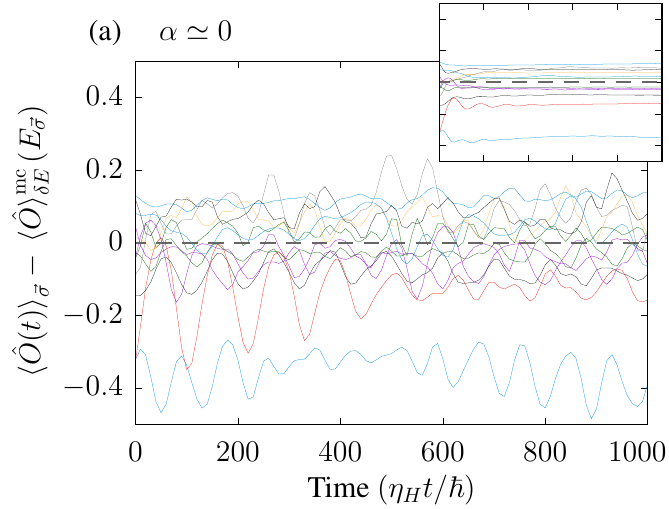}
    \includegraphics[width=0.48\linewidth]{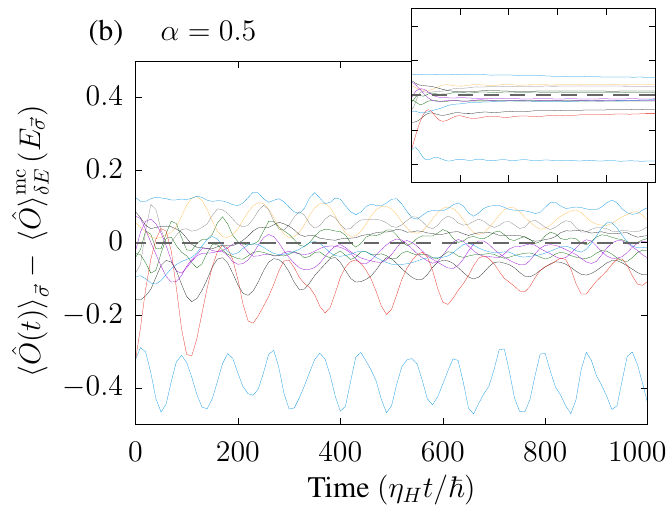}
    \includegraphics[width=0.48\linewidth]{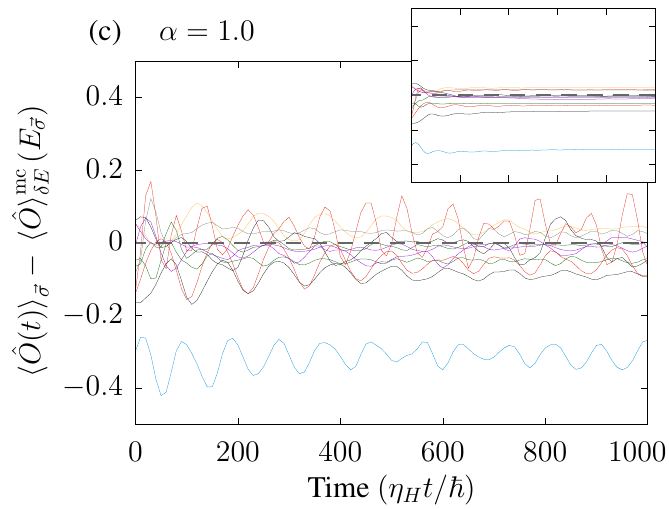}
    \includegraphics[width=0.48\linewidth]{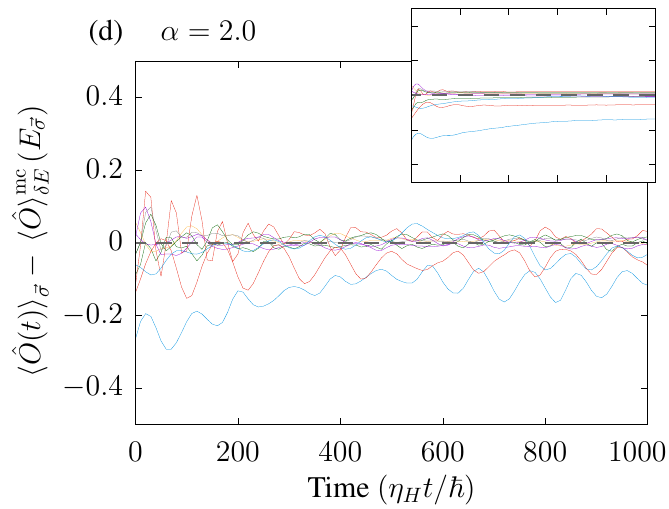}
    \includegraphics[width=0.48\linewidth]{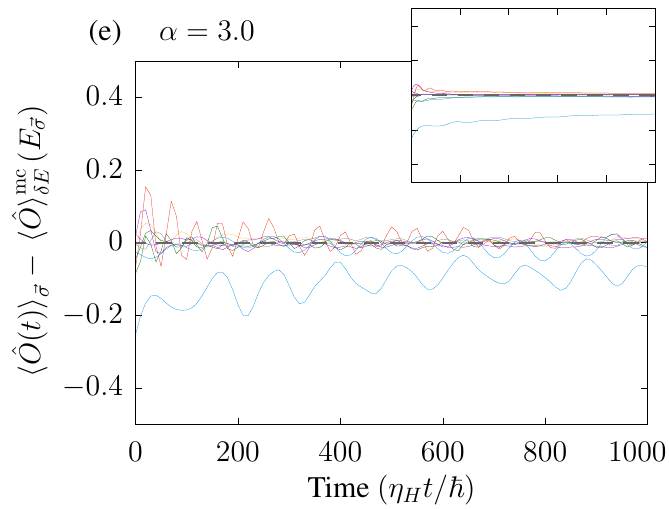}
    \includegraphics[width=0.48\linewidth]{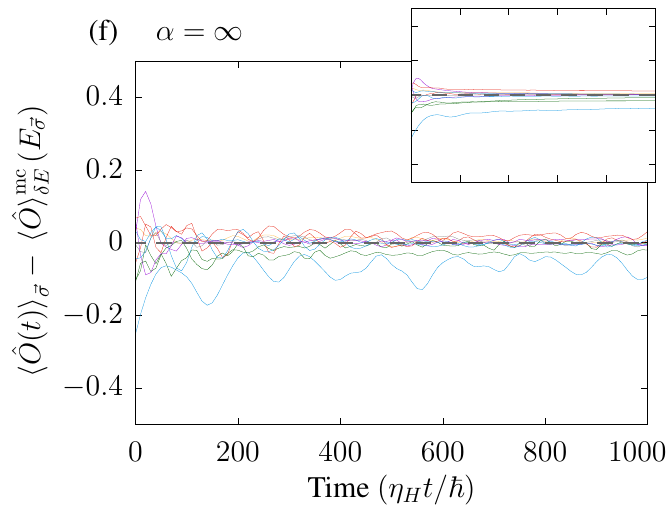}
    \caption{\label{fig:SMDynamics}
        Time evolution of the expectation value $\expval*{\hat{O}(t)}_{\vec{\sigma}}$ for several values of $\alpha$.
        Each inset shows the cumulative dynamics $\overline{\expval*{O}_{\vec{\sigma}}}(T)$ for the same data as in the main panel.
        As $\alpha$ increases, both the deviation  $\abs{\overline{\expval*{O}_{\vec{\sigma}}}(T) - \expval*{\hat{O}}^{\mathrm{mc}}_{\delta E}(E_{\vec{\sigma}})}$ and the temporal fluctuation of $\expval*{\hat{O}(t)}_{\vec{\sigma}}$ become small.
    }
\end{figure}

We examine the dynamics of the expectation value $\expval*{ \hat{O}(t) }_{\vec{\sigma}} \coloneqq \expval*{ \hat{O}(t) }{ \vec{\sigma}; 0,+ }$ and its cumulative counterpart $\overline{\expval*{O}_{\vec{\sigma}}}(T)$ defined by
\begin{equation}
    \overline{\expval*{O}_{\vec{\sigma}}}(T) \coloneqq \frac{1}{T} \int_{0}^{T} \dd{t'} \expval*{ \hat{O}(t) }{ \vec{\sigma}; 0,+ }.
\end{equation}
Figure~\ref{fig:SMDynamics} shows $\expval*{ \hat{O}(t) }_{\vec{\sigma}}$ in main panels~(a)-(f) and $\overline{\expval*{O}_{\vec{\sigma}}}(T)$ in insets for several samples.
Here, for each sample, we pick up an initial state $\ket*{\vec{\sigma};0,+}$ such that the cumulative expectation value $\overline{\expval*{O}_{\vec{\sigma}}}(T_{1})$ after relaxation (we choose $T_{1} = 10^{6}\,\hbar/\eta_{H}$ in the numerical calculation, which is sufficiently large compared with the experimentally relevant timescale) deviates most from the microcanonical average $\expval*{\hat{O}}^{\mathrm{mc}}_{\delta E}(E_{\vec{\sigma}})$ among those states that satisfy the condition~\eqref{Eq:SMEnergy}.
As the range of interactions becomes shorter, the maximum deviation of the expectation value after relaxation from the microcanonical average becomes smaller.
In addition, temporal fluctuations of $\expval*{ \hat{O}(t) }_{\vec{\sigma}}$ are typically larger for smaller $\alpha$, which indicate that the off-diagonal matrix elements $\matrixel*{ E_{\alpha} }{ \hat{O} }{ E_{\beta} }$ also become typically large for long-range interacting systems.

In Fig.~\ref{fig:SMDynamics_2}, we present the $N$-dependence of the ensemble average of the maximum deviation
\begin{equation}
    \max_{\vec{\sigma}}\, \abs{\Delta \overline{\expval*{O}_{\vec{\sigma}}} }\qc 
    \Delta \overline{\expval*{O}_{\vec{\sigma}}} \coloneqq \overline{\expval*{O}_{\vec{\sigma}}}(T_{1}) -\expval*{\hat{O}}^{\mathrm{mc}}_{\delta E}(E_{\vec{\sigma}}),
\end{equation}
where the maximum is taken over all $\vec{\sigma}$ that satisfy the condition~\eqref{Eq:SMEnergy}.

Overall, the ensemble average $\mathbb{E}_{N}[\max_{\vec{\sigma}} \abs{\Delta \overline{\expval*{O}_{\vec{\sigma}}} }]$ shows a similar $N$-dependence as $\mathbb{E}_{N}[\Delta_{\infty}]$ which is shown by dashed curves.
It decreases as $N$ increases for $\alpha\geq1.0$, while it increases with $N$ for $\alpha\simeq 0$.
For $\alpha=0.5$, $\mathbb{E}_{N}[\max_{\vec{\sigma}} \abs{\Delta \overline{\expval*{O}_{\vec{\sigma}}} }]$ seems slightly decreasing for $N\geq14$, but the number of samples is insufficient to draw a definite conclusion.
From these results for $\ket{\vec{\sigma};0,+}$, we also conjecture that initial states such as product states $\ket*{\sigma_{1}}\otimes\cdots\otimes\ket*{\sigma_{N}}$ fail to thermalize in the presence of long-range interactions even for relatively large system sizes.

\clearpage
{
\begin{figure}
\vspace{-1cm}
    \centering
    \includegraphics[width=0.7\linewidth]{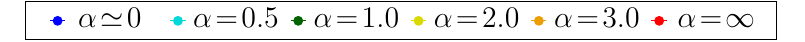}
    \includegraphics[width=0.48\linewidth]{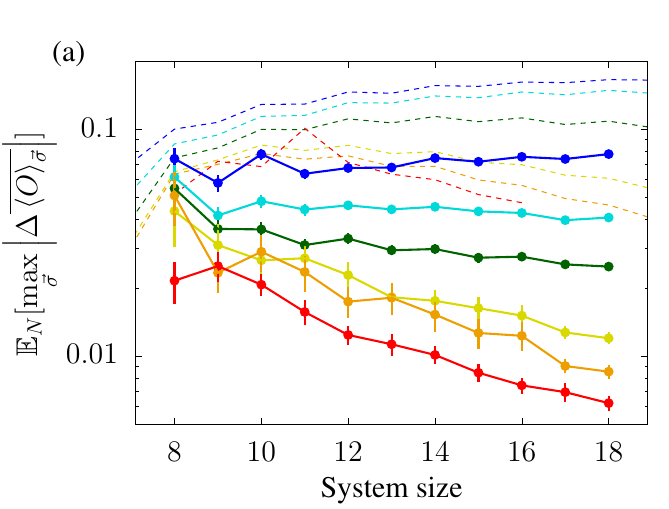}\\
    \includegraphics[width=0.48\linewidth]{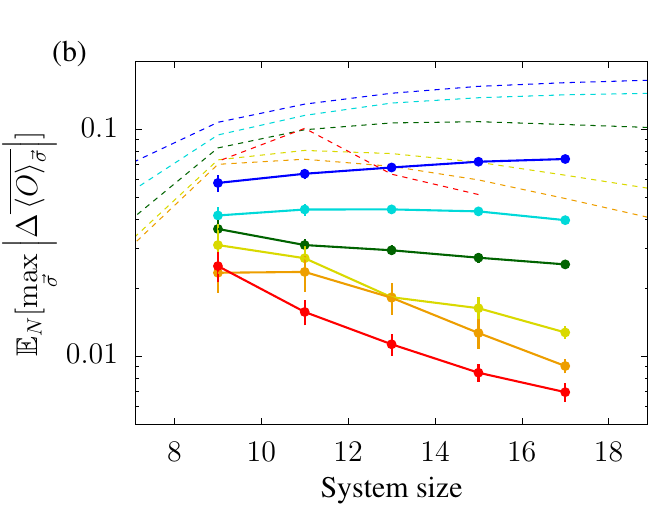}
    \includegraphics[width=0.48\linewidth]{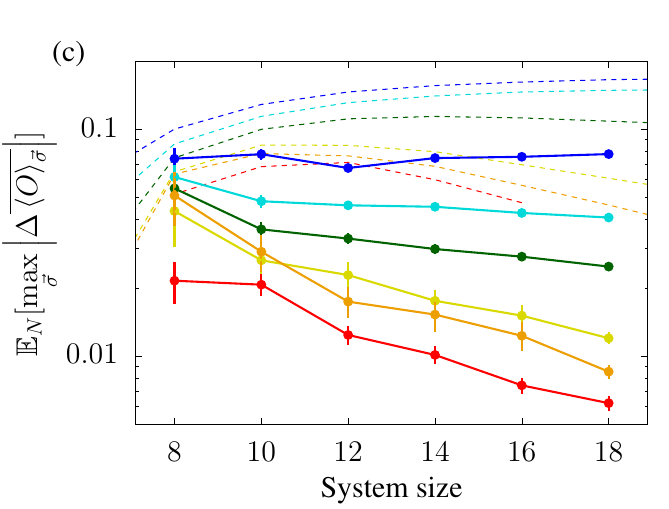}
    \caption{\label{fig:SMDynamics_2}
        Ensemble average of the maximum deviation $\max_{\vec{\sigma}} \abs{ \Delta \overline{\expval*{O}_{\vec{\sigma}}} }$ for (a) all $N$, (b) odd $N$, and (c) even $N$.
        Its $N$-dependence is similar to that of $\mathbb{E}_{N}[\Delta_{\infty}]$ shown by dashed curves.
        Here, $\mathbb{E}_{N}[\Delta_{\infty}]$ is calculated over the middle 20\% of the energy spectrum corresponding to the condition~\eqref{Eq:SMEnergy}.
        The number of samples are listed in Table~\ref{tab:SMNumberOfSamples} below.
    }
\end{figure}\nopagebreak\samepage
\begin{table}
    \centering
    \begin{tabular}{|c|c|c|c|c|c|c|c|c|} \hline
                 & 12      & 13       & 14       & 15       & 16       & 17      & 18      \\ \hline\hline
             0   & 333/997 & 358/997 & 404/1000 & 468/997  & 501/997  & 416/788 & 449/788 \\
             0.5 & 369/997 & 399/997 & 453/1000 & 514/1000 & 556/1000 & 517/880 & 552/880 \\
             1.0 & 361/996 & 407/997 & 462/997  & 531/1000 & 565/999  & 604/1000 & 641/1000 \\
             2.0 & 87/334  & 110/333 & 139/334  & 154/334  & 174/334  & 538/913 & 573/912 \\
             3.0 & 73/333  & 80/332  & 107/334  & 137/334  & 148/334  & 530/999 & 591/999 \\
        $\infty$ & 167/980 & 179/996 & 251/997  & 335/997  & 404/999  & 376/865 & 458/867 \\ \hline
    \end{tabular}
    \caption{\label{tab:SMNumberOfSamples}
        Number of samples in Fig.~\ref{fig:SMDynamics_2}.
        The number over the slash shows the number of samples for which $\vec{\sigma}$ satisfying the condition~\eqref{Eq:SMEnergy} exist and thus the quantity $\max_{\vec{\sigma}}\, \abs{\Delta \overline{\expval*{O}_{\vec{\sigma}}} }$ can be obtained.
        The number below the slash is the number of samples calculated in our numerical simulation.
    }
\end{table}
}

\clearpage
\section{Supplement to the section ``Range of validity of Srednicki's ansatz" in the main text}
\subsection{First part: typical magnitude of $\delta O_{\gamma\gamma}$}
In the main text, we introduce $\delta O_{\gamma\gamma} \coloneqq O_{\gamma\gamma} -\expval*{\hat{O}}^{\mathrm{mc}}_{\delta E}(E_{\gamma})$ and test the first part of Srednicki's ansatz that (i) $\mathcal{E}[\delta O_{\gamma\gamma}] = 0$ and $\mathcal{S}[\delta O_{\gamma\gamma}] = e^{ -\frac{ S(E_{\gamma}) }{2} } f(E_{\gamma})$.
We investigate the system-size dependence of the quantity
\begin{equation}
    \mathcal{S}^{E}_{\delta E} \coloneqq \sqrt{ \frac{1}{d_{E,\delta E}} \sum_{ \ket*{E_{\gamma}} \in \mathcal{H}_{E,\delta E} } \qty(\delta O_{\gamma\gamma} )^2 }
\end{equation}
for each sample, where $\mathcal{H}_{E,\delta E}$ is an energy shell centered at energy $E$ with a sufficiently small width $2\delta E$, and $d_{E,\delta E} \coloneqq \dim \mathcal{H}_{E,\delta E}$.
Srednicki's ansatz together with Boltzmann's formula implies that
$
    \mathcal{S}[\delta O_{\alpha\alpha}] \propto ( \sqrt{d_{E,\delta E}} )^{-a}.
$
Therefore, if Srednicki's ansatz typically holds, the distribution of $a$ over an ensemble should have a peak around unity.

\begin{figure}[tb]
    \centering
    \includegraphics[width=\linewidth]{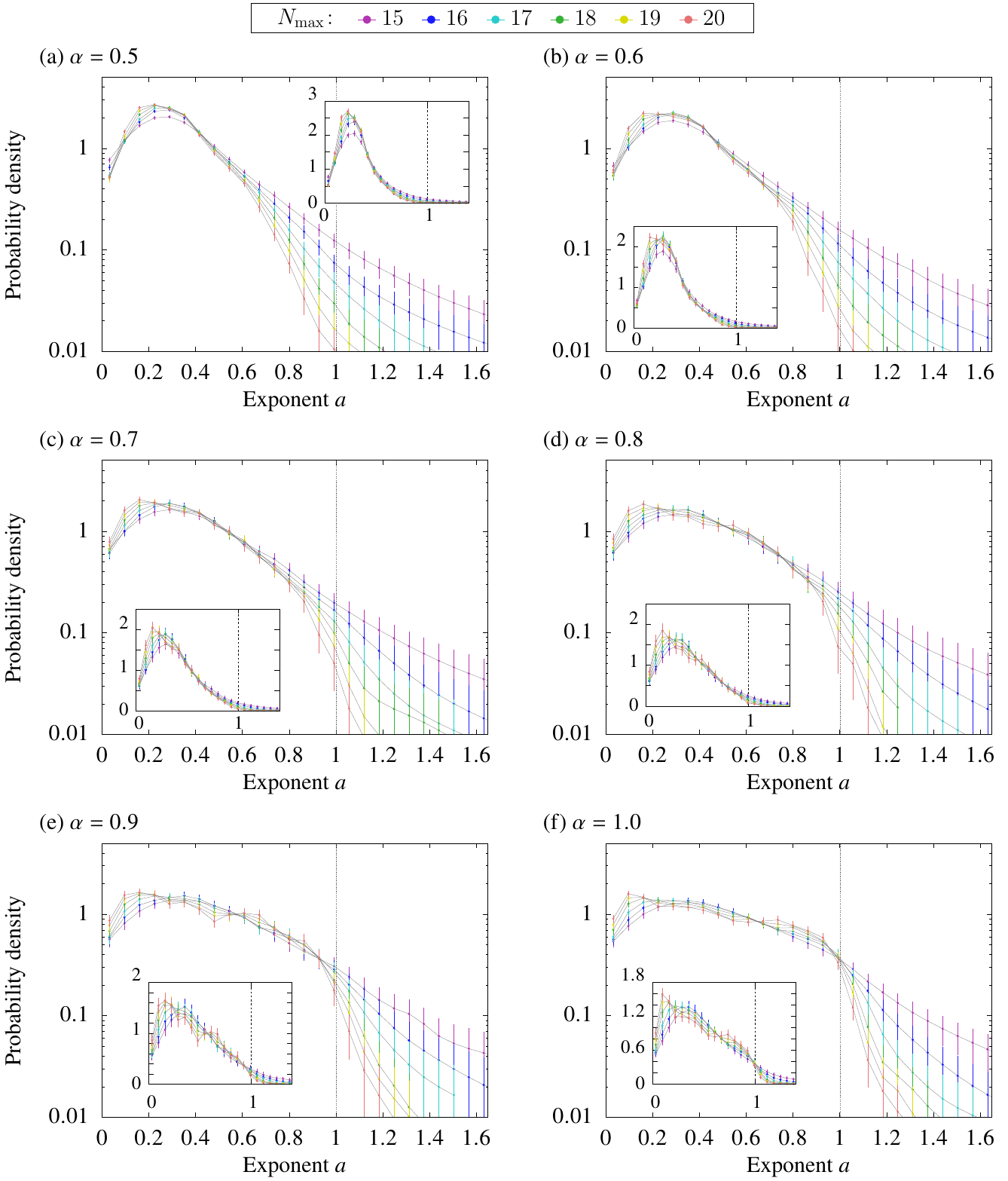}
    \caption{\label{fig:SMSrednickiAnsatz_1}
        Distribution of the exponent $a$ in the fitting $\mathcal{S}^{E}_{\delta E} \propto (\sqrt{d_{E,\delta E}})^{-a}$.
        Each inset shows the same graph as the main panel in the linear-linear scale.
        No peak around $a=1$ can be found for $\alpha\leq 1.0$, and the probability density around $a=1$ even decreases with increasing $N_{\max}$.
        This result indicates the breakdown of Srednicki's ansatz.
        The peak around $\alpha\simeq 0.2$ is an artifact of the fitting procedure, which always yields a positive value of $a$ even when $\mathcal{S}^{E}_{\delta E}$ decreases slower than an exponential function or does not decrease at all with increasing $d_{E,\delta E}$.
    }
\end{figure}

\begin{figure}[tb]
    \centering
    \includegraphics[width=\linewidth]{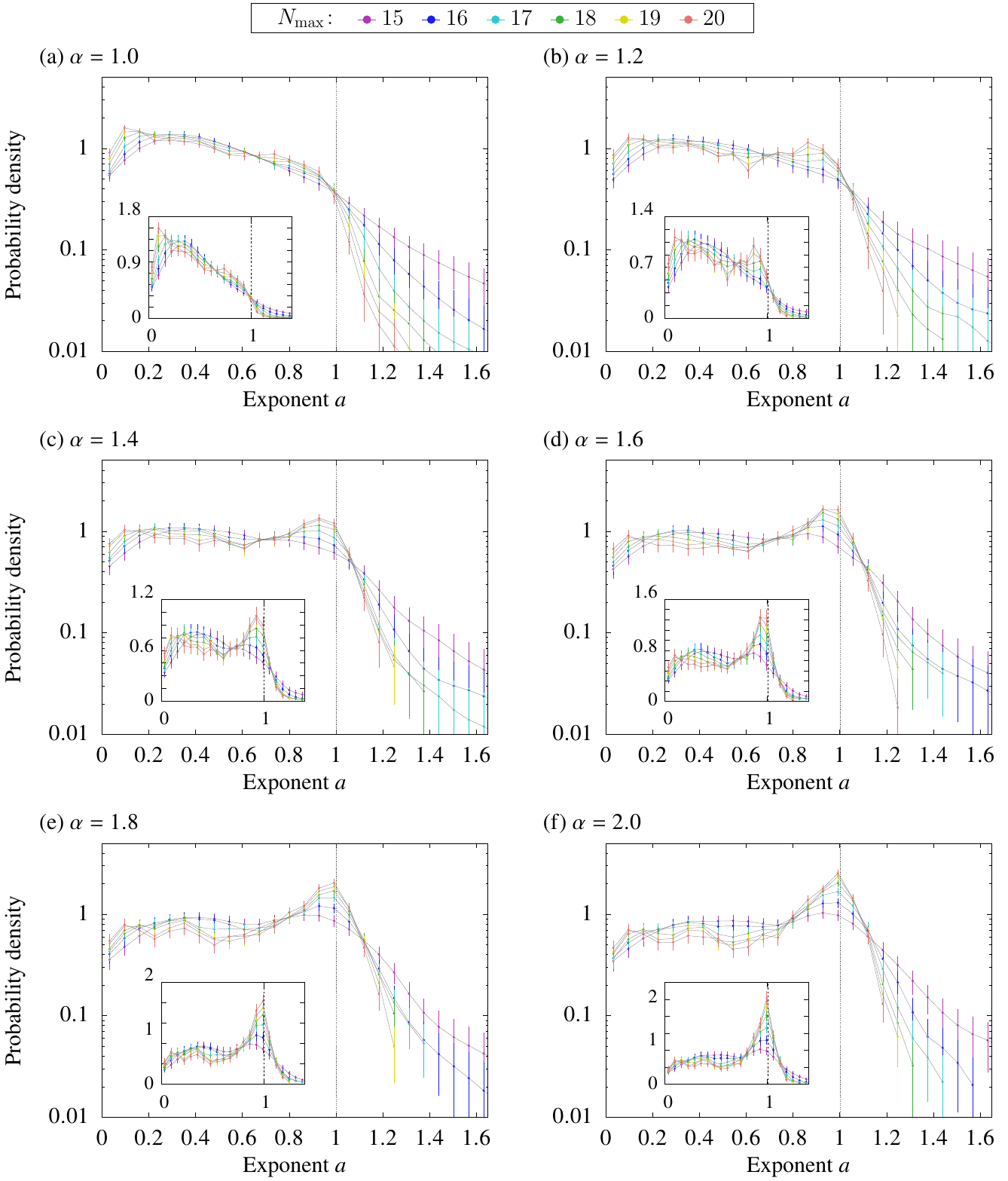}
    \caption{\label{fig:SMSrednickiAnsatz_2}
        Distribution of the exponent $a$ in the fitting $\mathcal{S}^{E}_{\delta E} \propto (\sqrt{d_{E,\delta E}})^{-a}$.
        Each inset shows the same graph as the main panel in the linear-linear scale.
        A peak appears around $a=1$ for $\alpha\geq 1.2$, and it develops as $N_{\max}$ increases, while the probability density for small $\alpha \:(\simeq 0.5)$ decreases with increasing $N_{\max}$.
        These results indicate that Srednicki's ansatz typically holds for $\alpha\geq 1.2$.
        Note that an increase in the probability density for $a \lesssim 0.2$ is an artifact of the fitting procedure, which always yields a positive value of $a$ even when $\mathcal{S}^{E}_{\delta E}$ decreases more slowly than an exponential function or does not decrease at all with increasing $d_{E,\delta E}$.
    }
\end{figure}

Figure~3 in the main text reports the results for $\alpha=0.0001,0.5,1.0$ and $3.0$.
Here, we present the data for intermediate values $\alpha=0.5,0.6,\dots,1.0$ in Fig.~\ref{fig:SMSrednickiAnsatz_1} and $\alpha=1.0,1.2,\dots,2.0$ in Fig.~\ref{fig:SMSrednickiAnsatz_2}.
For $\alpha\geq 1.2$, the distribution has a peak around unity, and the peak develops as the system size available for the fitting increases.
On the other hand, for $\alpha\leq 1.0$ there is no peak around unity, and the probability density around unity even decreases with increasing $N_{\max}$.
These results indicate that Srednicki's ansatz typically holds for $\alpha\geq 1.2$ but typically breaks down for $\alpha\leq 1.0$ at least for relatively large system size.

\clearpage
\subsection{Second part: distribution of $\delta O_{\gamma\gamma}$}
\begin{figure}
    \centering
    \includegraphics[width=0.48\linewidth]{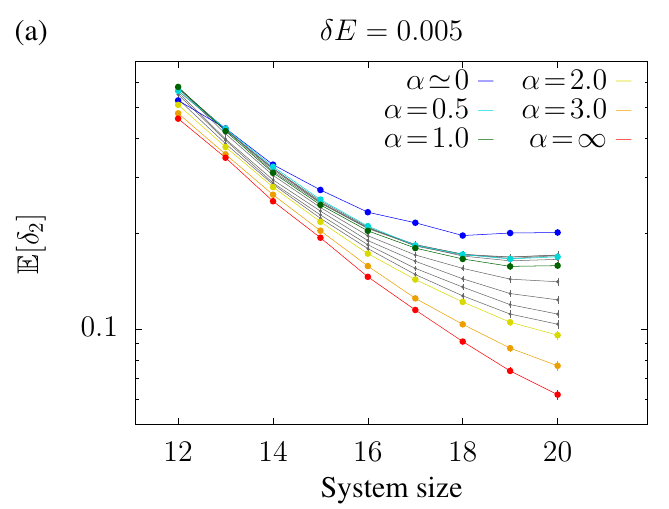}
    \includegraphics[width=0.48\linewidth]{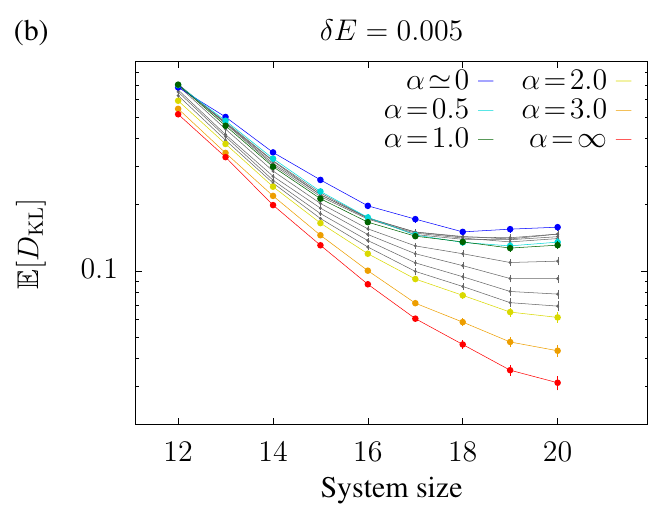}
    \includegraphics[width=0.48\linewidth]{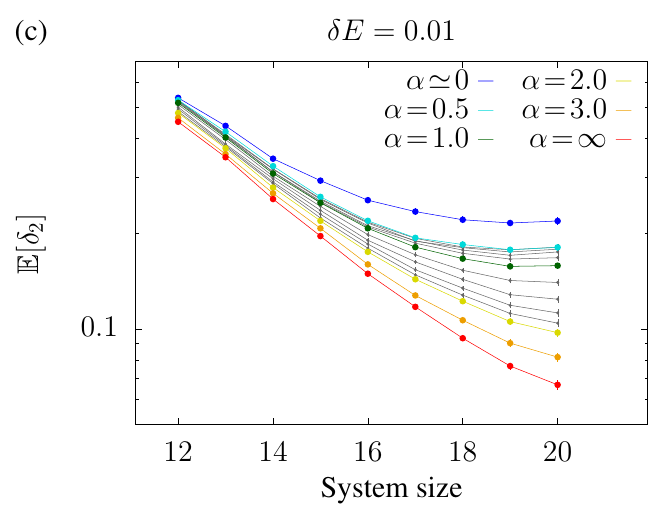}
    \includegraphics[width=0.48\linewidth]{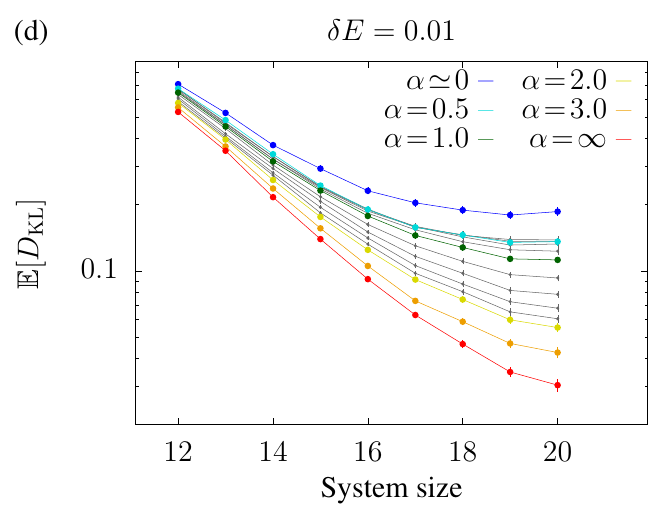}
    \includegraphics[width=0.48\linewidth]{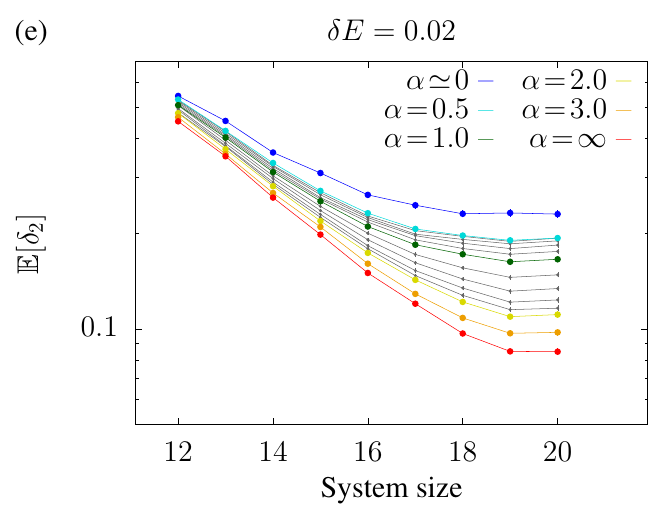}
    \includegraphics[width=0.48\linewidth]{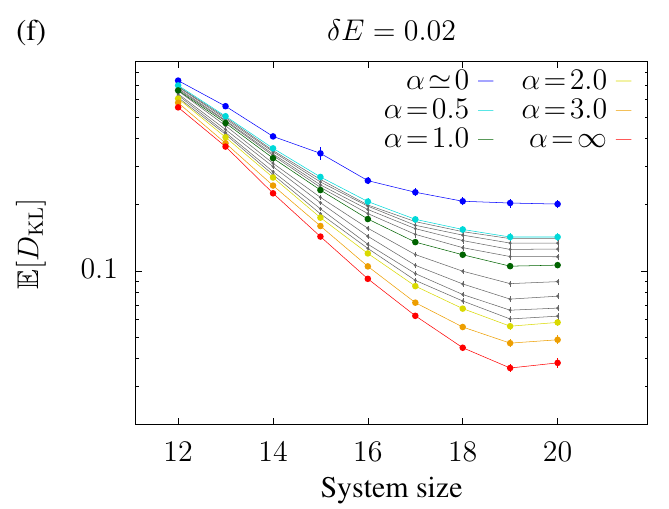}
    \caption{\label{fig:SMSrednickiAnsatz_distR}
        Ensemble average of (a)(c)(e) the L$2$-norm $\delta_{2}$ of the difference $P-P_{\mathrm{norm}}$ and (b)(d)(f) the KL-divergence of $P(R)$ relative to $P_{\mathrm{norm}}$, each for several values of the shell width $\delta E$.
        When the shell width is not small (i.e., $\delta E = 0.02$) and the energy dependence of $\expval*{\hat{O}}{E_{\alpha}}$ is not negligible in the calculation of $\mathcal{S}^{E}_{\delta E}$, both of $\mathbb{E}[\delta_{2}]$ and $\mathbb{E}[D_{\mathrm{KL}}]$ show convex behavior as seen for $\alpha=\infty$ in panels (e) and (f).
        For small $\delta E$ (i.e., $\delta E = 0.01, 0.005$), both $\mathbb{E}[\delta_{2}]$ and $\mathbb{E}[D_{\mathrm{KL}}]$ decrease with increasing $N$ for $\alpha\leq 2.0$ at least for the computationally available system size $(N\leq 20)$, which is consistent with Srednicki's ansatz.
        On the other hand, both $\mathbb{E}[\delta_{2}]$ and $\mathbb{E}[D_{\mathrm{KL}}]$ stop decreasing or even increase for large $N$ for $\alpha\lesssim 1.0$ irrespective of $\delta E$, indicating the breakdown of Srednicki's ansatz for long-range interacting systems.
    }
\end{figure}

We also test the second part of Srednicki's ansatz, i.e., (ii) $\delta O_{\gamma\gamma}$ behave like independent \textit{Gaussian} variables, by obtaining the distribution $\hat{P}_{N}$ of $\tilde{R}_{\gamma\gamma} \coloneqq \delta O_{\gamma\gamma} / \mathcal{S}^{E_{\gamma}}_{\delta E}$ for each sample $(\hat{H}_{N}^{(\alpha)}, \hat{O}_{N}^{(\infty)}) \in \mathcal{G}_{N}^{(\alpha)}\times\mathcal{G}_{N}^{(\infty)}$ and various $N$, which should be close to a normal distribution $P_{\mathrm{norm}}$ if Srednicki's ansatz holds and the shell width $d_{E,\delta E}$ is not too small.
We quantify the distance between $\hat{P}_{N}$ and $P_{\mathrm{norm}}$ in terms of the L$2$-norm 
\begin{align}
    \delta_{2} \coloneqq \bqty{ \Delta_{\mathrm{bin}} \sum_{\mathrm{bin}=1}^{ N_{\mathrm{bin}} } \qty( \hat{P}_{N}(R_{\mathrm{bin}}) - P_{\mathrm{norm}}(R_{\mathrm{bin}}) )^2 }^{1/2}, \label{eq:Deviation}
\end{align}
and the Kullback–Leibler divergence
\begin{equation}
    D_{\mathrm{KL}} \coloneqq \Delta_{\mathrm{bin}} \sum_{\mathrm{bin}=1}^{ N_{\mathrm{bin}} } \hat{P}_{N}(R_{\mathrm{bin}}) \log \frac{ \hat{P}_{N}(R_{\mathrm{bin}}) }{ P_{\mathrm{norm}}(R_{\mathrm{bin}}) },
\end{equation}
where $\Delta_{\mathrm{bin}} \, (=0.1)$ is the width of the bins in the calculation of the empirical probability density $\hat{P}(R)$, and $N_{\mathrm{bin}}$ is the number of bins.
Figure~\ref{fig:SMSrednickiAnsatz_distR} shows that the ensemble averages $\mathbb{E}_{N}[\delta_{2}]$ and $\mathbb{E}_{N}[D_{\mathrm{KL}}]$ decrease for $\alpha\gtrsim 1.0$ if we choose a sufficiently small energy shell $\delta E$, but they stop decreasing for large $N$ for $\alpha\lesssim 1.0$.
Therefore, the distribution of $\tilde{R}_{\alpha\alpha}$ deviates from a normal distribution at least for a relatively large system size, and Srednicki's ansatz breaks down also in this respect for long-range interacting systems with $\alpha\lesssim 1.0$.

\begin{figure}[b]
    \centering
    \includegraphics[width=0.48\linewidth]{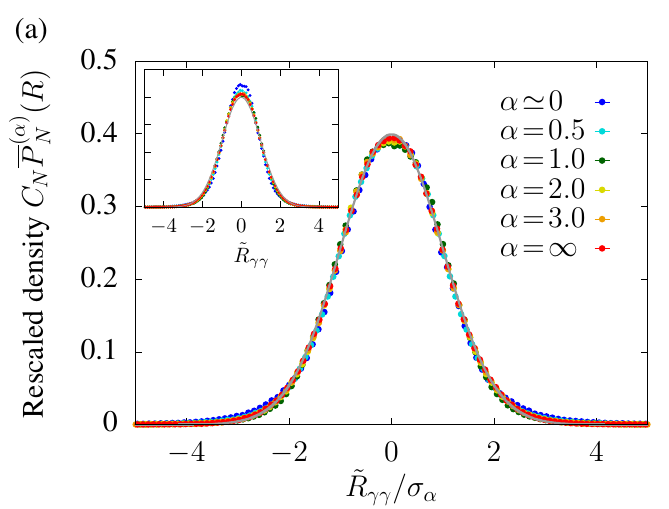}
    \includegraphics[width=0.48\linewidth]{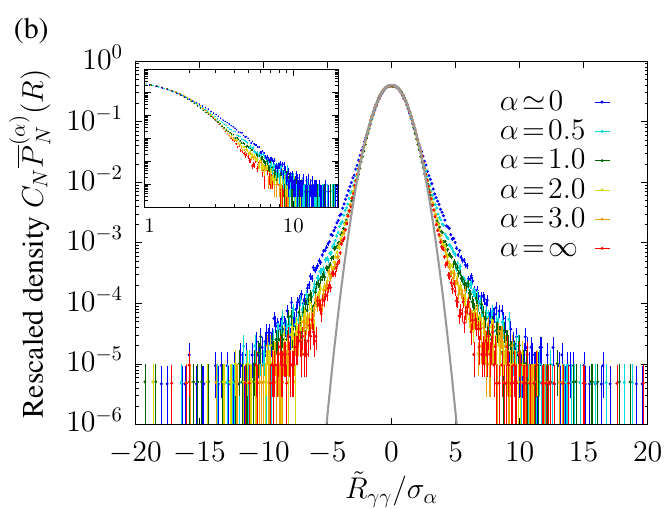}
    \caption{\label{fig:SMSrednickiAnsatz_distR1}
        Distribution of $\tilde{R}_{\gamma\gamma}$ over the whole ensembles with several decay exponents $\alpha$ rescaled by $\sigma_{\alpha}$ so that the distribution matches best to a normal distribution in the middle region $\abs*{R}\leq 2$ shown in (a) the linear-linear scale and (b) the log-linear scale.
        The inset in (a) shows the distribution $\overline{P}^{(\alpha)}(R)$ itself, while that in (b) shows $\tilde{R}_{\gamma\gamma}/\sigma_{\alpha}$ in the log-log scale.
        The y-axes of the insets are the same as that of the corresponding main panels.
        Grey curves show a normal distribution.
        The system size is $N=18$ for all the data.
        While $\overline{P}^{(\alpha)}(R)$ is well approximated by a Gaussian function in the middle $\abs*{R} \lesssim 2$, its tail is much heavier than that of a Gaussian distribution especially for small $\alpha$.
        As $\alpha$ increases, the tails become smaller and approach those of a Gaussian distribution.
    }
\end{figure}

We also calculate the distribution of $\tilde{R}_{\gamma\gamma}$ over each ensemble $\mathcal{G}^{(\alpha)}$, which we denote by $\overline{P}_{N}^{(\alpha)}(R)$, by collecting $\tilde{R}_{\gamma\gamma}$ for 1000 samples.
Figure~\ref{fig:SMSrednickiAnsatz_distR1} shows $\overline{P}_{N}^{(\alpha)}(R)$ for $N=18$ and several values of $\alpha$.
As $\alpha$ increases, the tails of $\overline{P}^{(\alpha)}(R)$ approach to those of a normal distribution.

\begin{figure}[bth]
    \centering
    \includegraphics[]{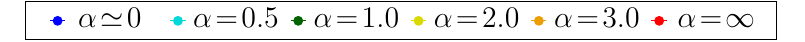}\\
    \includegraphics[width=0.48\linewidth]{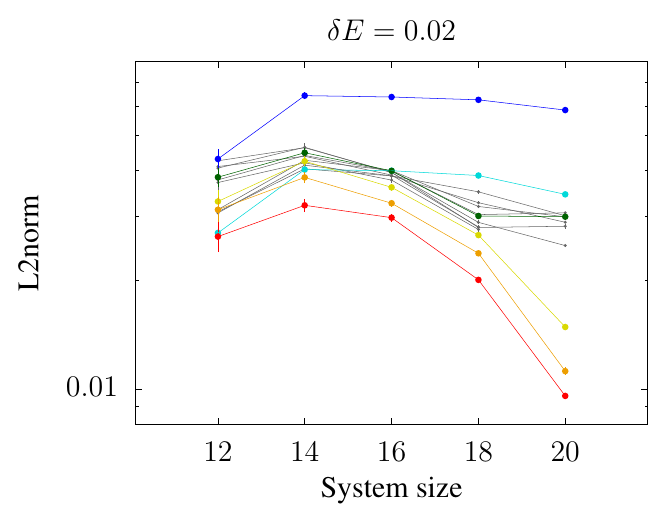}
    \includegraphics[width=0.48\linewidth]{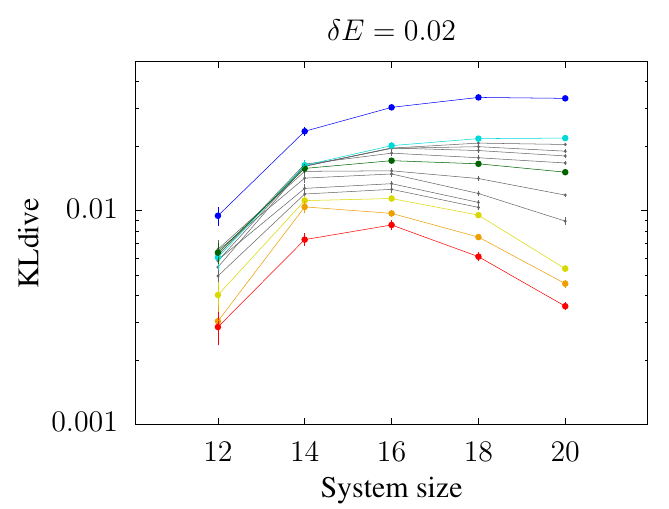}
    \includegraphics[width=0.48\linewidth]{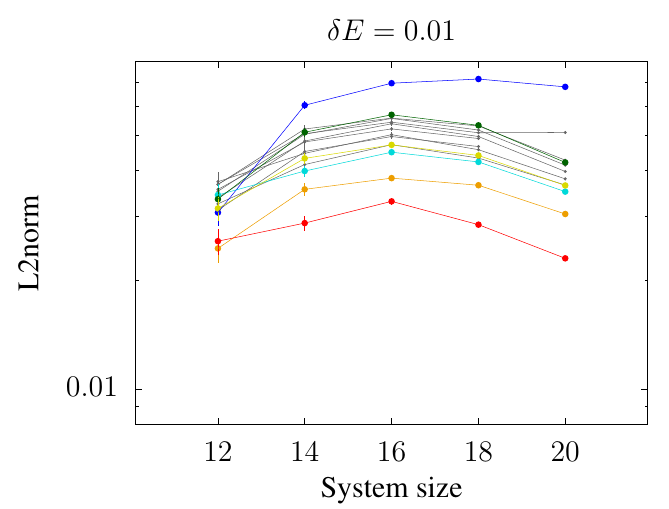}
    \includegraphics[width=0.48\linewidth]{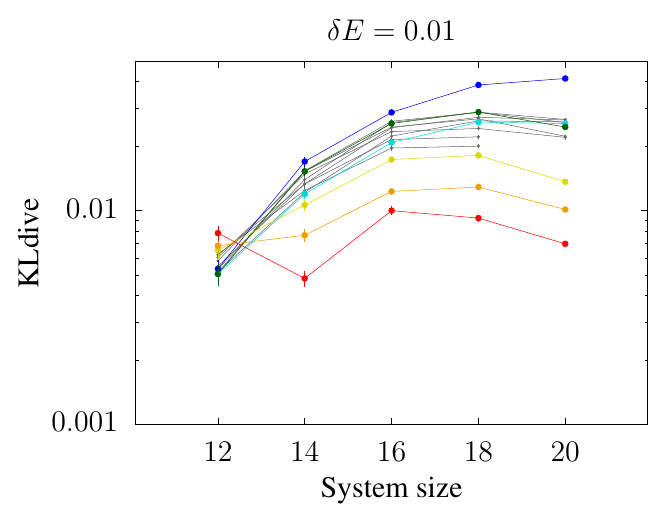}
    \includegraphics[width=0.48\linewidth]{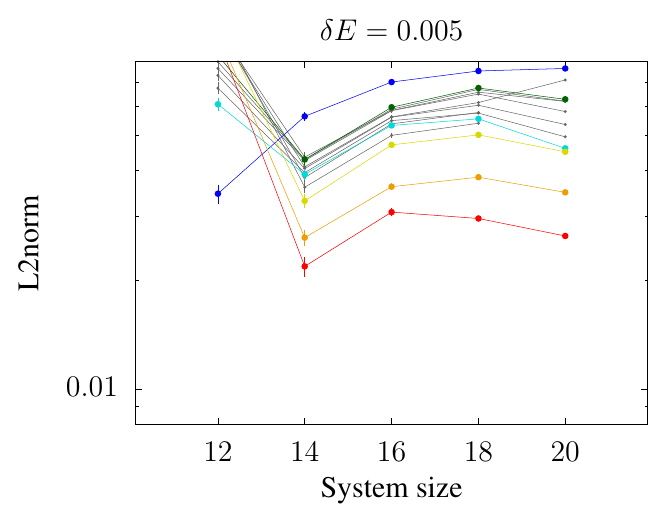}
    \includegraphics[width=0.48\linewidth]{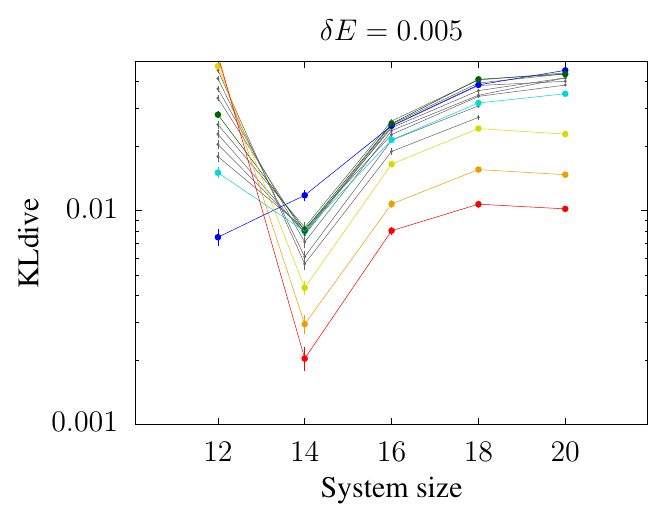}
    \caption{\label{fig:SMAverageDistribution_Measures}
        L$2$-norm of the difference $\overline{P}^{(\alpha)} - P_{\mathrm{norm}}$ and KL-divergence of $\overline{P}^{(\alpha)}$ relative to the normal distribution $P_{\mathrm{norm}}$ for several $\delta E$ used in the calculation of $\mathcal{S}^{E}_{\delta E}$.
        For $\alpha\geq2.0$, both of $\mathbb{E}[\delta_{2}]$ and $\mathbb{E}[D_{\mathrm{KL}}]$ decrease for large $N$ irrespective of the choice of $\delta E$.
        For $\alpha<2.0$, however, the $N$-dependence of these quantities depends on the value of $\delta E$, which prevents us from drawing a conclusion about whether $\overline{P}^{(\alpha)}$ approaches a normal distribution or not.
    }
\end{figure}

Figure~\ref{fig:SMAverageDistribution_Measures} shows the $N$-dependence of the L$2$-norm $\overline{\delta}_{2}$ and the Kullback–Leibler divergence $\overline{D}_{\mathrm{KL}}$ for $\overline{P}^{(\alpha)}(R)$ for several choices of the shell width $\delta E$ used in the calculation of $\mathcal{S}^{E}_{\delta E}$.
For $\alpha\geq2.0$, both $\overline{\delta}_{2}$ and $\overline{D}_{\mathrm{KL}}$ decrease for large $N$ irrespective of $\delta E$, which is consistent with Srednicki's ansatz.
For $\alpha<2.0$, however, the $N$-dependence of these quantities depends on the value of $\delta E$, which prevents us from deciding whether $\overline{P}^{(\alpha)}$ approaches a normal distribution or not.

\begin{figure}[htb]
    \vspace{-1cm}
    \centering
    \includegraphics[width=0.48\linewidth]{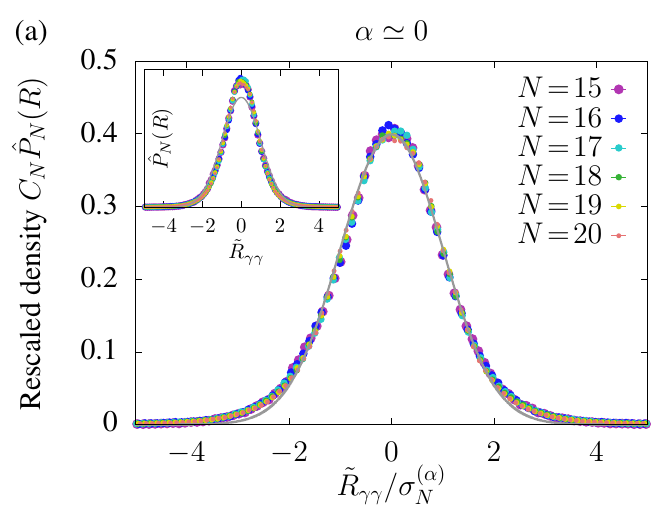}
    \includegraphics[width=0.48\linewidth]{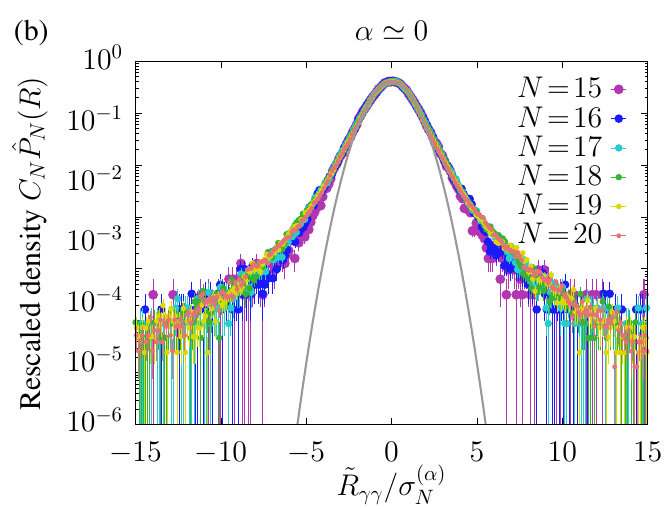}
    \includegraphics[width=0.48\linewidth]{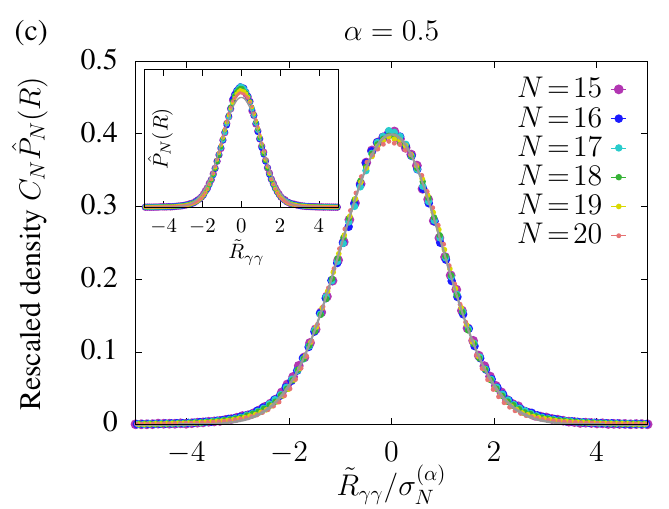}
    \includegraphics[width=0.48\linewidth]{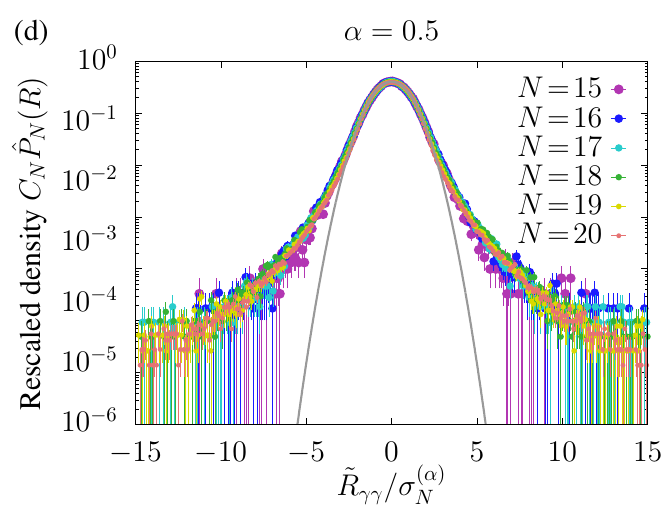}
    \includegraphics[width=0.48\linewidth]{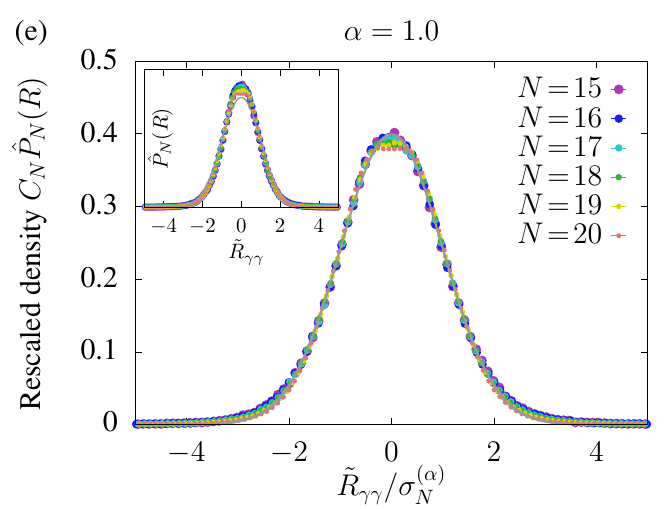}
    \includegraphics[width=0.48\linewidth]{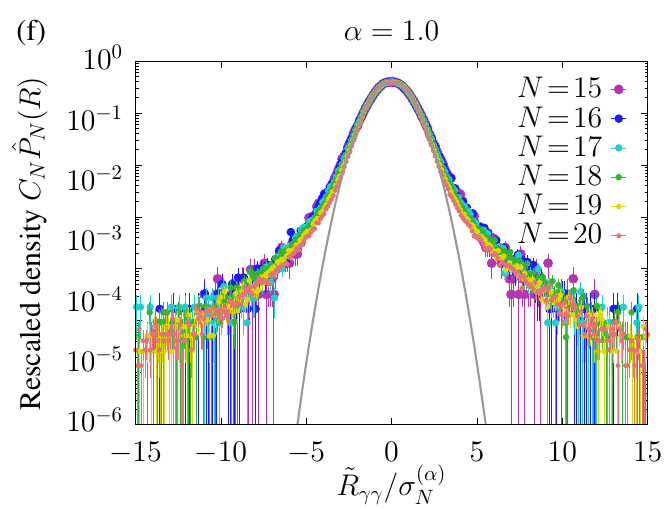}
    \caption{\label{fig:SMSrednickiAnsatz_distR2}
    Distribution of $\tilde{R}_{\alpha\alpha} \coloneqq \delta O_{\alpha\alpha} / \mathcal{S}^{E_{\alpha}}_{\delta E}$ over the whole ensemble for $\alpha\leq 1.0$ and several $N$ rescaled so that it matches best to a normal distribution in the middle region ($\abs*{R}\leq 2$).
    Grey curves show normal distributions.
    The tails of the distribution do not seem to approach those of a normal distribution.
    }
\end{figure}

\begin{figure}
    \centering
    \includegraphics[width=0.48\linewidth]{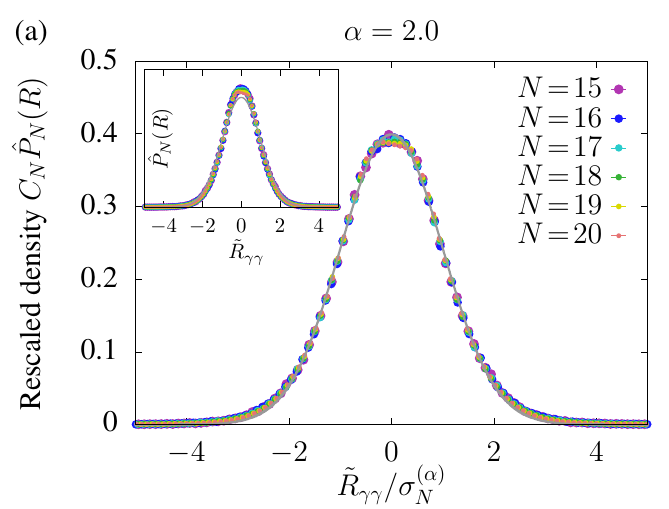}
    \includegraphics[width=0.48\linewidth]{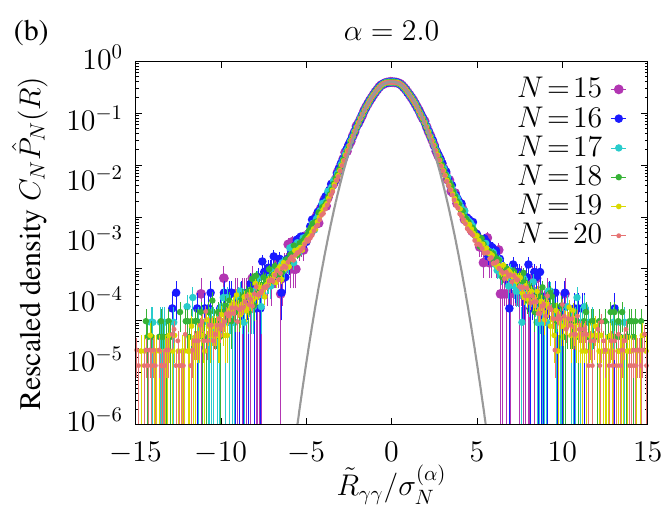}
    \includegraphics[width=0.48\linewidth]{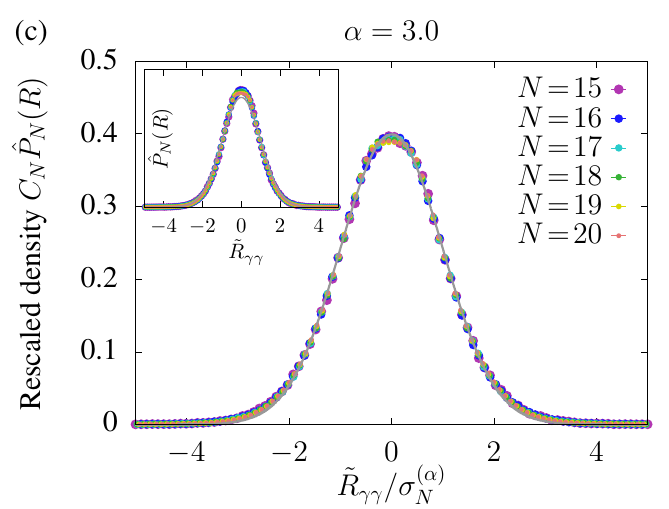}
    \includegraphics[width=0.48\linewidth]{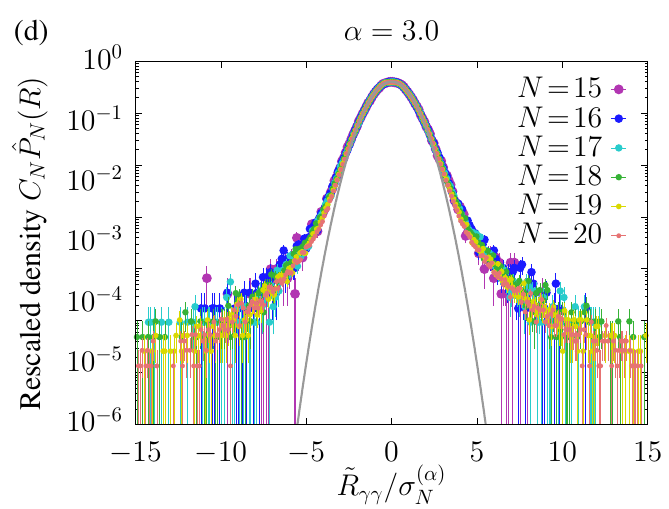}
    \includegraphics[width=0.48\linewidth]{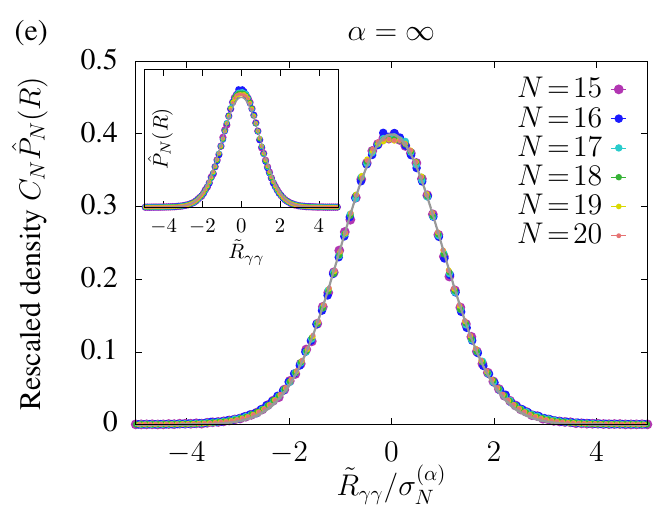}
    \includegraphics[width=0.48\linewidth]{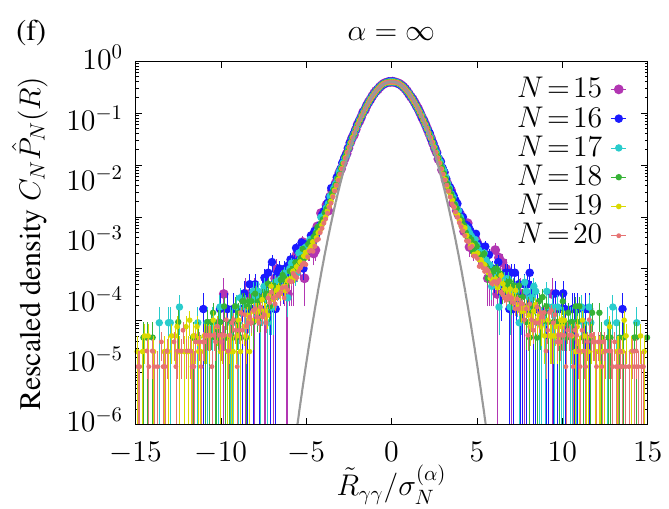}
    \caption{\label{fig:SMSrednickiAnsatz_distR3}
    Distribution of $\tilde{R}_{\alpha\alpha} \coloneqq \delta O_{\alpha\alpha} / \mathcal{S}^{E_{\alpha}}_{\delta E}$ over the whole ensemble for $\alpha\geq 2.0$ and several $N$ rescaled so that it matches best to a normal distribution in the middle region ($\abs*{R}\leq 2$).
    Grey curves show normal distributions.
    The distribution seems to approach a normal distribution as $N$ increases.
    }
\end{figure}

Figures~\ref{fig:SMSrednickiAnsatz_distR2} and \ref{fig:SMSrednickiAnsatz_distR3} show $\overline{P}_{N}^{(\alpha)}(R)$ itself for several values of $\alpha$ and system sizes $N$.
Although the dependence on the shell width $\delta E$ cannot be neglected\footnote{A large $\delta E$ fails to eliminate the contribution from the energy dependence of $\expval*{\hat{O}}^{\mathrm{mc}}_{\delta E}(E_{\alpha})$, while small $\delta E$ suffers from an insufficient number of states in an energy shell $\mathcal{H}_{E_{\alpha},\delta E}$ leading to large finite-size effects.},
it seems that the tails of $\overline{P}^{(\alpha)}$ for $\alpha\lesssim 1.0$ do not approach those of a normal distribution even when $N$ becomes large, while
the distribution for $\alpha\geq 2.0$ seems to approach to a normal distribution as the system size $N$ increases.

\clearpage
\section{Permutation symmetries and $\mathfrak{su}(d_{L})$ algebra in the fully connected case}
In the main text, we argue that the strong ETH breaks down for a fully connected Hamiltonian $\hat{H}^{(\alpha=0)}$ because of the permutation symmetry between any two sites.
The permutation symmetry implies that we can block-diagonalize $\hat{H}^{(0)}$ according to irreducible representations of a permutation group.
On the other hand, $\hat{H}^{(0)}$ can also be block-diagonalized according to irreducible representations of the $\mathfrak{su}(d_{L})$ algebra, where $d_{L}$ is the dimension of the local Hilbert space on each site.
For $d_{L}=2$, it is known that these two ways of block-diagonalization give the equivalent result~\citeSM{bapst2012quantum}.

In this section, we prove that this equivalence also holds for arbitrary $d_{L}$.
From the representation theory of operator algebra, we can decompose the total Hilbert space $\mathcal{H}_{\mathrm{loc}}^{\otimes N}$ by the symmetric group $\mathfrak{S}_{N}$ among $N$ sites as
\begin{equation}
    \mathcal{H}_{\mathrm{loc}}^{\otimes N} = \bigoplus_{q}\mathcal{H}_{\mu_{q}}\otimes \mathcal{H}_{\nu_{q}},
\end{equation}
where $\mathcal{H}_{\mu_{q}}$ is an irreducible representation of the group algebra $\mathbb{C}\mathfrak{S}_{N}$, and $\dim \mathcal{H}_{\nu_{q}}$ gives its multiplicity.
Correspondingly, we have
\begin{equation}
    \mathbb{C}\mathfrak{S}_{N} \simeq \bigoplus_{q} \mathcal{L}(\mathcal{H}_{\mu_{q}})\otimes \hat{I}_{\nu_{q}}, \qq{(Wedderburn decomposition~\citeSM{wedderburn1908hypercomplex,kabernik2021reductions}),}
\end{equation}
where $\mathcal{L}(\mathcal{H}_{\mu_{q}})$ denotes the space of all operators acting on $\mathcal{H}_{\mu_{q}}$, and $\hat{I}_{\nu_{q}}$ denotes the identity operator on $\mathcal{H}_{\nu_{q}}$.
Moreover, the commutant of the symmetric group $\mathfrak{S}_{L}$ in $\mathcal{L}(\mathcal{H}_{\mathrm{loc}}^{\otimes L} )$ defined by
\begin{equation}
    \mathrm{comm}( \mathfrak{S}_{L} ) \coloneqq \Bqty{ \hat{A} \in \mathcal{L}(\mathcal{H}_{\mathrm{loc}}^{\otimes L} ) \mid \forall \hat{\tau} \in \mathfrak{S}_{L},\ \comm*{\hat{A}}{\hat{\tau}} = 0 },
\end{equation}
is isomorphic to $\bigoplus_{q} \hat{I}_{\mu_{q}}\otimes\mathcal{L}( \mathcal{H}_{\nu_{q}})$.
Therefore, in order to show the equivalence between the block diagonalization by the permutation symmetry of any two sites and that by the $\mathfrak{su}(d_{L})$ algebra generated by $\Bqty{ \hat{M}^{(q)} \coloneqq \sum_{j=1}^{N} \hat{\sigma}^{(q)}_{j} }_{q=1}^{d_{L}}$, it is sufficient to prove that 
\begin{equation}
    \mathfrak{su}(d_{L}) \simeq \mathrm{comm}( \mathfrak{S}_{L} ).
\end{equation}

Since every $\hat{M}^{(q)}$ is invariant under any permutation $\hat{\tau} \in \mathfrak{S}_{L}$, it is clear that $\mathfrak{su}(d_{L}) \subseteq \mathrm{comm}( \mathfrak{S}_{L} )$.
We now prove the other direction $\mathrm{comm}( \mathfrak{S}_{L} ) \subseteq \mathfrak{su}(d_{L})$.
Let $\hat{A}$ be an operator that commutes with all permutations, i.e., $\comm*{\hat{A}}{\hat{\tau}} = 0$ for all $\hat{\tau}\in\mathfrak{S}_{L}$, and expand $\hat{A}$ in terms of direct products of a local operator basis $\Bqty*{ \hat{\sigma}^{p} }$ as
\begin{equation}
    \hat{A} = \sum_{n=1}^{L} \sum_{p_{1}\dots p_{n}} \sum_{j_{1}<j_{2}<\dots<j_{n}} c^{(n)}_{\vec{j},\vec{p}} \ \hat{\sigma}^{p_{1}}_{j_{1}}\cdots \hat{\sigma}^{p_{n}}_{j_{n}}.
\end{equation}
Then, the commutation relation $\comm*{\hat{A}}{\hat{\tau}} = 0 \ (\forall\hat{\tau} \in \mathfrak{S}_{L})$ gives
\begin{align}
    \hat{A} 
    &= \frac{1}{ L! } \sum_{ \tau\in\mathfrak{S}_{L} } \hat{\tau}^{-1} \hat{A} \hat{\tau}
    = \sum_{n=1}^{L} \sum_{p_{1}\dots p_{n}} \sum_{j_{1}<j_{2}<\dots<j_{n}} c^{(n)}_{\vec{j},\vec{p}} \ \qty( \frac{1}{ L! } \sum_{ \tau\in\mathfrak{S}_{L} } \hat{\sigma}^{p_{1}}_{\tau(j_{1})}\dots \hat{\sigma}^{p_{n}}_{\tau(j_{n})} ). \label{Eq:SMcommutant}
\end{align}
By choosing a permutation $\pi_{\vec{j}}$ such that $\pi_{\vec{j}}(\alpha) = j_{\alpha}$ for $\alpha=1,\dots,n$, we can rewrite the above equation as
\begin{align}
    \hat{A} 
    &= \sum_{n=1}^{L} \sum_{p_{1}\dots p_{n}} \sum_{j_{1}<j_{2}<\dots<j_{n}} c^{(n)}_{\vec{j},\vec{p}} \ \qty( \frac{1}{ L! } \sum_{ \tau\in\mathfrak{S}_{L} } \hat{\sigma}^{p_{1}}_{\tau\circ\pi_{\vec{j}}(1)}\dots \hat{\sigma}^{p_{n}}_{\tau\circ\pi_{\vec{j}}(n)} ) \nonumber \\
    &= \sum_{n=1}^{L} \sum_{p_{1}\dots p_{n}} \qty(\sum_{j_{1}<j_{2}<\dots<j_{n}} c^{(n)}_{\vec{j},\vec{p}} ) \ \frac{1}{ L! } \qty( \sum_{ \tau\in\mathfrak{S}_{L} } \hat{\sigma}^{p_{1}}_{\tau(1)}\dots \hat{\sigma}^{p_{n}}_{\tau(n)} ).
\end{align}

To show that $\hat{A}$ belongs to $\mathfrak{su}(d_L)$, it is sufficient to prove the relation
\begin{equation}
    \sum_{ \tau\in\mathfrak{S}_{L} } \hat{\sigma}^{p_{1}}_{\tau(1)}\dots \hat{\sigma}^{p_{n}}_{\tau(n)} \in \mathfrak{su}(d_{L}) \label{Eq:SMIndHypothesis}
\end{equation}
by induction.
For $n=1$, it is easy to see that
\begin{equation}
    \sum_{ \tau\in\mathfrak{S}_{L} } \hat{\sigma}^{p_{1}}_{\tau(1)} = (L-1)!\, \hat{M}^{p_{1}} \in \mathfrak{su}(d_{L})
\end{equation}
for any $p_{1}$.
Assume that Eq.~\eqref{Eq:SMIndHypothesis} holds for all $n< N$, and consider the relation
\begin{align}
    \hat{M}^{p_{1}}\dots \hat{M}^{p_{N}} 
    &= \sum_{j_{1}\dots j_{N}} \hat{\sigma}^{p_{1}}_{j_{1}}\dots \hat{\sigma}^{p_{N}}_{j_{N}} \nonumber \\
    &= \sum_{ \substack{j_{1}\dots j_{N} \\ \forall \alpha\neq\beta,\ j_{\alpha} \neq j_{\beta} } } \hat{\sigma}^{p_{1}}_{j_{1}}\dots \hat{\sigma}^{p_{N}}_{j_{N}} +(\text{terms with $n< N$}) \nonumber \\
    &= \sum_{ \tau\in\mathfrak{S}_{L} } \hat{\sigma}^{p_{1}}_{\tau(1)}\dots \hat{\sigma}^{p_{N}}_{\tau(N)} +(\text{terms with $n< N$}).
\end{align}
Since any $\hat{M}^{p}$ commutes with all permutations, the same calculation as in Eq.~\eqref{Eq:SMcommutant} implies that the last term in the above equation is a linear combination of the terms of the form
\begin{equation}
    \sum_{ \tau\in\mathfrak{S}_{L} } \hat{\sigma}^{p_{1}}_{\tau(1)}\dots \hat{\sigma}^{p_{n}}_{\tau(n)}
\end{equation}
with $n<N$, which are assumed to belong to $\mathfrak{su}(d_{L})$ by the induction hypothesis.
Therefore, we obtain
\begin{equation}
    \sum_{ \tau\in\mathfrak{S}_{L} } \hat{\sigma}^{p_{1}}_{\tau(1)}\dots \hat{\sigma}^{p_{N}}_{\tau(N)} = \hat{M}^{p_{1}}\hat{M}^{p_{2}}\cdots \hat{M}^{p_{N}} + (\text{terms with $n< N$}) \in \mathfrak{su}(d_{L}),
\end{equation}
which completes the proof.

\clearpage
\bibliographystyleSM{apsrev4-2}
\bibliographySM{supplement}

\end{document}